\documentclass[
 aps,prd,reprint,amsmath,amssymb,
 superscriptaddress,nofootinbib,floatfix,
]{revtex4-2}
\usepackage{graphicx}\usepackage{dcolumn}\usepackage{bm}
\usepackage{hyperref}\usepackage{xcolor}

\begin{document}

\title{Observational properties of magnetized quantum Oppenheimer--Snyder black holes}

\author{Anuar Idrissov}
\email{anuar.idrissov@gmail.com}
\affiliation{Instituto de Ciencias Nucleares, Universidad Nacional Aut\'onoma de M\'exico, Mexico}
\affiliation{Fesenkov Astrophysical Institute, Observatory 23, 050020, Almaty, Kazakhstan}
\affiliation{Al-Farabi Kazakh National University, Al-Farabi Ave.\ 71, 050040, Almaty, Kazakhstan}

\author{Hernando~\surname{Quevedo}}
\email[]{quevedo@nucleares.unam.mx}
\affiliation{Instituto de Ciencias Nucleares, Universidad Nacional Aut\'onoma de M\'exico, Mexico}
\affiliation{Dipartimento di Fisica and ICRA, Universit\`a di Roma “La Sapienza”, Roma, Italy}
\affiliation{Al-Farabi Kazakh National University, Al-Farabi Ave.\ 71, 050040, Almaty, Kazakhstan}
\date{\today}

\begin{abstract}
We consider a thin accretion disk around the quantum Oppenheimer--Snyder black
hole immersed in an asymptotically uniform magnetic field, together with the
gravitational capture of test particles by the black hole. The field is
described as a Wald test field and acts only on the weakly charged matter of
the disk. We find that the quantum correction affects the disk only slightly,
while the magnetic coupling strongly increases its radiative efficiency and
luminosity and suppresses the Doppler blue peak of the iron K$\alpha$ line,
which becomes a single peak below the rest frame energy. On the other hand, the
quantum correction reduces all capture cross sections, while the magnetic field
does not affect the capture of photons and confines charged particles to a
finite region. Therefore, in principle, the two effects may be distinguished by
combining spectral and shadow observations.
\end{abstract}

\maketitle

\section{Introduction}\label{sec:intro}

It is well known that in General Relativity (GR) the complete gravitational
collapse of a massive star leads to the formation of a black hole with a
space-time singularity at its center~\cite{Penrose1965,HawkingPenrose1970}. The
simplest example is the Oppenheimer--Snyder (OS) model of a homogeneous dust
cloud~\cite{OS1939}, for which the singularity is reached in a finite co-moving
time. Since singularities are generally regarded as a signal of the breakdown
of the theory, one expects that quantum effects will become important in the
last stages of collapse and that they may resolve the singularity (see for
example~\cite{Malafarina2017} for a review). Following this idea, Lewandowski
\textit{et al.}~\cite{Lewandowski2023} considered the OS model with corrections
inspired by loop quantum cosmology. The resulting quantum Oppenheimer--Snyder
(qOS) collapse avoids the classical singularity, while the exterior geometry is
static, spherically symmetric and depends on a single additional parameter
$\alpha$. Notice however that the exterior metric is not regular at $r=0$. This
is not a problem in the collapse model, since the metric describes the
space-time only outside the collapsing cloud. Interestingly, an
exterior with the same asymptotic behavior was obtained from dust collapse in
asymptotic safety~\cite{BMP2024}, where the corrections come from the running of
the gravitational coupling. The qOS exterior has been studied in several
contexts, such as its thermodynamics and topological properties~\cite{Dong2025},
its shadow~\cite{Luo2024}, quasinormal modes~\cite{Skvortsova2024}, greybody
factors~\cite{Lv2025}, particle dynamics and quasiperiodic
oscillations~\cite{Bouzenada2025}, tidal forces~\cite{2026arXiv260903278I},
higher-dimensional and charged extensions~\cite{Shi2024,Mazharimousavi2025},
tunneling and entropy~\cite{Tan2025}, and constraints from extreme mass-ratio
inspirals~\cite{Yang2026}.

Since horizons can not be observed directly, the possibility of distinguishing
such quantum corrected black holes from the classical Schwarzschild solution
relies on the observation of their surroundings. Accretion disks are probably
the most important example. In the Novikov--Thorne (NT)
model~\cite{NovikovThorne1973,PageThorne1974} the disk is geometrically thin and
optically thick, it radiates locally as a black body, and its flux, temperature
and luminosity are entirely determined by the circular orbits of the background
geometry. Therefore even small departures from Schwarzschild can leave an imprint
on the disk spectrum~\cite{ShakuraSunyaev1973,Bambi2013}. This idea has been
applied to a large number of compact objects, for example black holes surrounded
by dark matter with different pressure
profiles~\cite{Boshkayev2020,Kurmanov2022,Boshkayev2022}, space-times with a
quadrupole moment and rotation~\cite{Boshkayev2021,Boshkayev2024HT}, rotating
regular black holes and black holes in non-linear
electrodynamics~\cite{Boshkayev2024reg,Kurmanov2024}, rotating naked
singularities~\cite{Kurmanov2025}, black holes in Hernquist~\cite{Nieto2025} and
King~\cite{Zare2025} dark matter halos and wormholes~\cite{HarkoKovacsLobo2009}.

Another important probe of the inner part of the disk is the iron K$\alpha$
line. The disk is illuminated by a hot corona and the reflected spectrum
contains a fluorescent line at $6.4$~keV, whose observed shape is distorted by
the motion of the gas and by the gravitational field close to the compact
object~\cite{Fabian1989,Tanaka1995,Laor1991}. The line has been used to measure
black hole spins~\cite{Reynolds2014} and to test the Kerr
hypothesis~\cite{Bambi2017,JohannsenPsaltis2013}, as well as to study exotic
compact objects and regular black holes~\cite{BambiMalafarina2013,Abdikamalov2019,Urmanov2025}.

A complementary probe is given by the gravitational capture cross section.
While the disk observables depend on the geometry close to the innermost stable
circular orbit, the capture cross section is determined by the peak of the
effective potential, i.e. by the photon sphere in the case of massless
particles, and therefore it probes a different region of the
space-time~\cite{Zakharov1994,Shapiro1983,FrolovNovikov1998}. Capture cross
sections have been computed for the Reissner--Nordstr\"om
geometry~\cite{Zakharov1994}, for Schwarzschild--Tangherlini black
holes~\cite{Ahmedov2021}, for the Zipoy--Voorhees
($q$-metric)~\cite{Momynov2025} and
Bocharova--Bronnikov--Melnikov--Bekenstein space-times~\cite{Turimov2025}, and
more recently for charged black holes in scalar--tensor--vector
gravity~\cite{Sucu2025}. For the marginally bound orbit and the massless and
massive cross sections we shall follow the treatment of the last reference.

Finally, astrophysical compact objects are generally immersed in magnetic
fields. Even when such fields are too weak to modify the geometry, they can
strongly affect the motion of charged
matter~\cite{AlievGaltsov1989,FrolovShoom2010,Kolos2015}. The standard
description of a uniform field around a stationary and axially symmetric black
hole is given by the Wald solution~\cite{Wald1974}, where the electromagnetic
potential is constructed from the Killing vectors of the space-time.

In the present article we combine the above ingredients and study the
accretion disk and the capture of particles for the qOS black hole immersed in
a uniform magnetic field. The paper is organized as follows. In
Sec.~\ref{sec:bg} we review the qOS space-time, discuss the allowed range of
$\alpha$ and introduce the magnetic field. In Sec.~\ref{sec:orbits} we obtain
the charged circular orbits and the innermost stable circular orbit (ISCO).
Section~\ref{sec:disk} is devoted to the radiative properties of the disk, while
in Sec.~\ref{sec:iron} we compute the iron K$\alpha$ line. In
Sec.~\ref{sec:capture} we discuss the capture cross sections and finally in
Sec.~\ref{sec:concl} we summarize the results. Throughout the article we use
units with $G=c=1$ and, in the numerical results, we set $M=1$. Unless
otherwise stated, in all figures the left panel shows the case without magnetic
field for $\alpha=-1,\,0,\,1,\,2,\,2.8$, while the right
panel shows the effect of the magnetic coupling for
$|b|=0,\,0.05,\,0.10,\,0.15,\,0.20$ at fixed $\alpha=0.6$.

\section{The space-time}\label{sec:bg}

\subsection{The qOS metric and the range of $\alpha$}

The exterior of the qOS black hole is described by the line element
\begin{equation}
ds^2=-f(r)\,dt^2+\frac{dr^2}{f(r)}+h(r)\left(d\theta^2+\sin^2\theta\,d\phi^2\right),
\label{eq:metric}
\end{equation}
with
\begin{equation}
f(r)=1-\frac{2M}{r}+\frac{\alpha M^2}{r^4},\qquad h(r)=r^2 ,
\label{eq:f}
\end{equation}
where $\alpha=16\sqrt{3}\pi\gamma^{3}\ell_{\mathrm{Pl}}^{2}$, $\gamma$ is the
Barbero--Immirzi parameter and $\ell_{\rm Pl}$ is the Planck
length.\footnote{The parameter $\alpha$ has dimensions of $M^2$, and in the
numerical results it is given in units of $M^2$, so that for example $\alpha=2.8$
stands for $\alpha=2.8M^2$.} The metric is asymptotically flat but it is not Ricci
flat, since
$R^\mu{}_\nu=\left(\alpha M^2/r^6\right)\,{\rm diag}(-6,-6,3,3)$ and the Ricci
scalar is $R=-6\alpha M^2/r^6$. The Kretschmann scalar is
\begin{equation}
K=\frac{12M^2}{r^{12}}\left(4r^6-20\alpha Mr^3+39\alpha^2M^2\right),
\label{eq:kretschmann}
\end{equation}
which reduces to the Schwarzschild value $48M^2/r^6$ for $\alpha=0$.

Loop quantum cosmology suggests $\alpha\ge0$. However the metric is well
defined also for negative values and it is instructive to consider the whole
range of $\alpha$ for which a thin disk can exist. We can identify three
regimes. For $\alpha<0$ the correction term makes the gravitational potential
deeper and there is a single horizon outside $2M$ (for example $r_h=2.107M$ for
$\alpha=-1$). For $0<\alpha<27/16$ there are two horizons, which merge at
\begin{equation}
\alpha_{\rm hor}=\frac{27}{16}=1.6875,\qquad r_h=\frac{3M}{2}.
\label{eq:alphahor}
\end{equation}
For $\alpha>\alpha_{\rm hor}$ the object is horizonless. Nevertheless it still
has a photon sphere and an ISCO, so the disk model is still well defined. The
real upper limit is set by the photon sphere. As we shall see in
Sec.~\ref{sec:capture}, the photon sphere is given by the largest root of
$r^4-3Mr^3+3\alpha M^2=0$, which disappears when the quartic has a double root at
$r=9M/4$, that is for
\begin{equation}
\alpha_{\max}=\frac{729}{256}=2.847656 .
\label{eq:alphamax}
\end{equation}
For $\alpha>\alpha_{\max}$ there is no photon sphere and therefore no critical
impact parameter, no shadow and no photon capture. The ISCO, however,
exists for larger values, since the condition~\eqref{eq:iscopoly} has a
root up to $\alpha_{\rm ISCO}^{\max}=5.353234$, at
$r=\tfrac38\left(5\sqrt{13}-7\right)M=4.1354M$, where the minimum and the
maximum of $E(r)$ coincide. However, for $\alpha>\alpha_{\max}$ the timelike
circular orbits extend inwards all the way to the static ring
$r=(2\alpha M)^{1/3}$, where $f'=0$ and $L=0$. Then there is no effective
capture surface replacing the horizon and the zero torque condition at the
inner edge of the NT disk loses its meaning. For this reason we shall restrict
the attention to $-1\le\alpha\le2.8$, which includes all three regimes and stays
just below $\alpha_{\max}$. For $|\alpha|\ll M^2$ the horizon and the other
characteristic radii can also be written as power series in $\alpha$, which we
report in Appendix~\ref{app:series}.

\subsection{External uniform magnetic field}

The metric~\eqref{eq:metric} has two Killing vectors,
$\xi_{(t)}^\mu=\delta^\mu_t$ and $\xi_{(\phi)}^\mu=\delta^\mu_\phi$. Following
Wald~\cite{Wald1974}, a test field describing an asymptotically uniform magnetic
field $B$ aligned with the symmetry axis can be written as
$A^\mu=\tfrac{B}{2}\xi^\mu_{(\phi)}+C\,\xi^\mu_{(t)}$. For a rotating black hole
the constant $C$ is fixed by requiring that the black hole is not charged and it
is proportional to the spin. Since the qOS geometry is static ($g_{t\phi}=0$)
this term vanishes and we are left with
\begin{equation}
A_t=0,\qquad A_\phi=\frac{B}{2}\,h(r)\sin^2\theta=\frac{B}{2}r^2\sin^2\theta,
\label{eq:Amu}
\end{equation}
with $A_r=A_\theta=0$.

Two remarks are in order here. First, the field is a test field and its
backreaction on the geometry is negligible for
$B\ll B_M\simeq2.35\times10^{19}(M_\odot/M)$~G, which is the case for all
astrophysical values. Second, Wald's construction relies on the identity
$\Box\xi^\mu=-R^\mu{}_{\nu}\xi^\nu$, which turns a Killing vector into a vacuum
solution of Maxwell's equations only when the space-time is Ricci flat. This is
not the case for the qOS metric and therefore the potential~\eqref{eq:Amu} must
be understood as a test field sustained by an effective current
$j^\mu=\nabla_\nu F^{\mu\nu}=2R^\mu{}_{\nu}A^\nu$. Using computer
algebra~\cite{Bonanos} we find that the only non vanishing component is
\begin{equation}
\nabla_\nu F^{\phi\nu}=\frac{3\alpha BM^2}{r^6}=-\frac{B}{2}\,R ,
\label{eq:maxwellres}
\end{equation}
which vanishes in the Schwarzschild limit. Notice that we do not perform any
expansion in $\alpha$ in the following.

Of course a test field does not affect neutral matter. Therefore we shall
assume that the disk is made of weakly charged matter with charge-to-mass ratio
$\bar q=q/m$, so that all the magnetic effects are described by the coupling
$b\equiv\bar q B$. In the equations only the dimensionless combination $bM$
appears and all the values of $|b|$ quoted below are values of $|b|M$.

\section{Charged circular orbits}\label{sec:orbits}

\begin{figure*}[t]\includegraphics[width=0.98\textwidth]{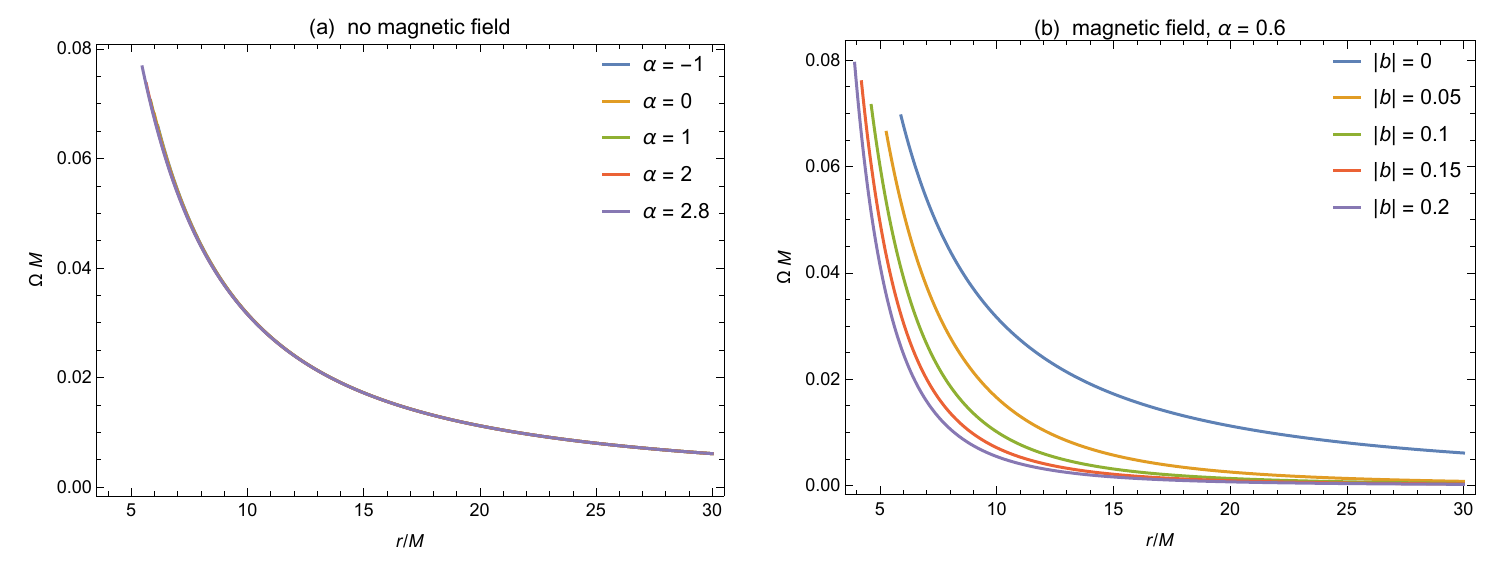}
\caption{Orbital angular velocity $\Omega M$ of the charged circular orbits.
Left: no magnetic field, for different values of $\alpha$. Right: different
values of the magnetic coupling $|b|=|\bar qB|M$ for $\alpha=0.6$. Each curve
starts at the corresponding ISCO.}\label{fig:omega}
\end{figure*}

\subsection{Effective potential and circular orbits}

The motion of a charged test particle is described by the Lagrangian
\begin{equation}
    \mathcal{L}=\tfrac12 g_{\mu\nu}\dot x^\mu\dot x^\nu+\bar q A_\mu\dot x^\mu,
\end{equation}
where the dot denotes the derivative with respect to proper time. The two
Killing vectors give the conserved specific energy and angular momentum
\begin{equation}
E=f\,\dot t,\qquad L=r^2\sin^2\theta\,\dot\phi+\tfrac{b}{2}r^2\sin^2\theta .
\end{equation}
Restricting to the equatorial plane, the normalization
$g_{\mu\nu}\dot x^\mu\dot x^\nu=-1$ gives $\dot r^2=E^2-V_{\rm eff}$, with the
effective potential
\begin{equation}
V_{\rm eff}(r)=f(r)\left[1+\frac{\left(L-\tfrac{b}{2}r^2\right)^2}{r^2}\right].
\label{eq:Veff}
\end{equation}
Circular orbits are given by $V_{\rm eff}=E^2$ and $V_{\rm eff}'=0$. If we
define $X\equiv L-\tfrac b2 r^2$ the second condition becomes a quadratic
equation for $X$,
\begin{equation}
\left(f'r-2f\right)X^2-2fb\,r^2X+f'r^3=0,
\label{eq:quad}
\end{equation}
and then we obtain
\begin{equation}
E=\sqrt{f\Big(1+\frac{X^2}{r^2}\Big)},\quad
\Omega=\frac{fX}{r^2E},\quad
L=X+\frac{b}{2}r^2 .
\label{eq:ELO}
\end{equation}
Of the two roots of Eq.~\eqref{eq:quad} we take the one with $X>0$, which
reduces to the prograde Schwarzschild orbit for $b\to0$. Notice that the
canonical angular momentum $L$ grows without bound at large radii, while the
kinetic part decays as $X\simeq M/(br)$, so that $E\to1$ and the binding energy
is well defined. Eqs.~\eqref{eq:quad} and \eqref{eq:ELO} are invariant under
$(b,L)\to(-b,-L)$, and therefore the disk observables depend only on $|b|$.

\subsection{Innermost stable circular orbit}

The inner edge of the disk is located at the ISCO, $r_{\rm ISCO}$, which
corresponds to the minimum of $E(r)$~\cite{BPT1972}. For $\alpha=b=0$ we recover
the Schwarzschild values $r_{\rm ISCO}=6M$, $E_{\rm ISCO}=2\sqrt2/3$,
$L_{\rm ISCO}=2\sqrt3 M$ and $\Omega_{\rm ISCO}=(6\sqrt6M)^{-1}$. The values for the
other cases are reported in Table~\ref{tab:res}.

For $b=0$ the ISCO can be found from a polynomial equation. In fact, circular
geodesics in a static space-time are marginally stable when
\begin{equation}
    3ff'+rff''-2r{f'}^2=0,
\end{equation}
which for the lapse~\eqref{eq:f} becomes
\begin{equation}
r^7-6Mr^6+4\alpha Mr^4+9\alpha M^2r^3-12\alpha^2M^3=0 ,
\label{eq:iscopoly}
\end{equation}
and gives $r_{\rm ISCO}=6M$ for $\alpha=0$. For $b\neq0$ there is no such
simplification, because $X(r)$ is itself a root of Eq.~\eqref{eq:quad}. If we
eliminate $X$ between $dE/dr=0$ and Eq.~\eqref{eq:quad} we obtain a resultant
whose relevant factor is of degree twenty in $r$ and quadratic in $b^2$.
Therefore in the magnetized case we obtain $r_{\rm ISCO}(\alpha,b)$ numerically
from the minimum of $E(r)$. On the other hand, for $b=0$ and small $\alpha$ the
root of Eq.~\eqref{eq:iscopoly} can be expanded as
$r_{\rm ISCO}\simeq6M-11\alpha/(72M)$ (see Appendix~\ref{app:series} for the
higher orders).

\begin{table*}[t]
\caption{Disk and capture quantities for the magnetized qOS black hole
($M=1$, $\dot m=1$, $\sigma=1$). The upper block shows different values of
$\alpha$ at $|b|=0$, the lower block different values of the magnetic coupling
at $\alpha=0.6$. $r_{\mathcal{F}}$ is the radius at which the flux is maximum,
while the last two columns give the radius and the angular momentum of the
marginally bound orbit. The first row of the lower block is the field free
case at $\alpha=0.6$, which we use as a reference for the magnetic ratios quoted
in the text.}
\label{tab:res}
\begin{ruledtabular}
\begin{tabular}{cccccccccc}
$\alpha$ & $|b|$ & $r_{\rm ISCO}/M$ & $E_{\rm ISCO}$ & $\eta\,(\%)$
 & $\mathcal{F}_{\max}(\times10^{5})$ & $r_{\mathcal{F}}/M$
 & $T_{\max}(\times10^{2})$ & $r_{\rm mb}/M$ & $L_{\rm mb}/M$\\
\hline
$-1.0$ & 0.00 & 6.1449 & 0.9438 & 5.617  & 1.288  & 9.77 & 5.990  & 4.1166 & 4.0296 \\
 0.0   & 0.00 & 6.0000 & 0.9428 & 5.719  & 1.368  & 9.55 & 6.081  & 4.0000 & 4.0000 \\
 1.0   & 0.00 & 5.8377 & 0.9416 & 5.836  & 1.464  & 9.31 & 6.186  & 3.8640 & 3.9667 \\
 2.0   & 0.00 & 5.6512 & 0.9403 & 5.973  & 1.583  & 9.04 & 6.308  & 3.6969 & 3.9280 \\
 2.8   & 0.00 & 5.4767 & 0.9390 & 6.102  & 1.702  & 8.79 & 6.423  & 3.5220 & 3.8911 \\
\hline
 0.6   & 0.00 & 5.9050 & 0.9421 & 5.787  & 1.423  & 9.41 & 6.142  & 3.9213 & 3.9805 \\
 0.6   & 0.05 & 5.2762 & 0.8801 & 11.989 & 4.106  & 8.73 & 8.005  & 3.4571 & 4.3144 \\
 0.6   & 0.10 & 4.6304 & 0.8396 & 16.042 & 9.777  & 7.27 & 9.944  & 3.2342 & 4.5926 \\
 0.6   & 0.15 & 4.2104 & 0.8085 & 19.155 & 17.022 & 6.35 & 11.422 & 3.0915 & 4.8421 \\
 0.6   & 0.20 & 3.9189 & 0.7827 & 21.728 & 25.459 & 5.73 & 12.632 & 2.9887 & 5.0728 \\
\end{tabular}
\end{ruledtabular}
\end{table*}

\begin{figure*}[t]\includegraphics[width=0.98\textwidth]{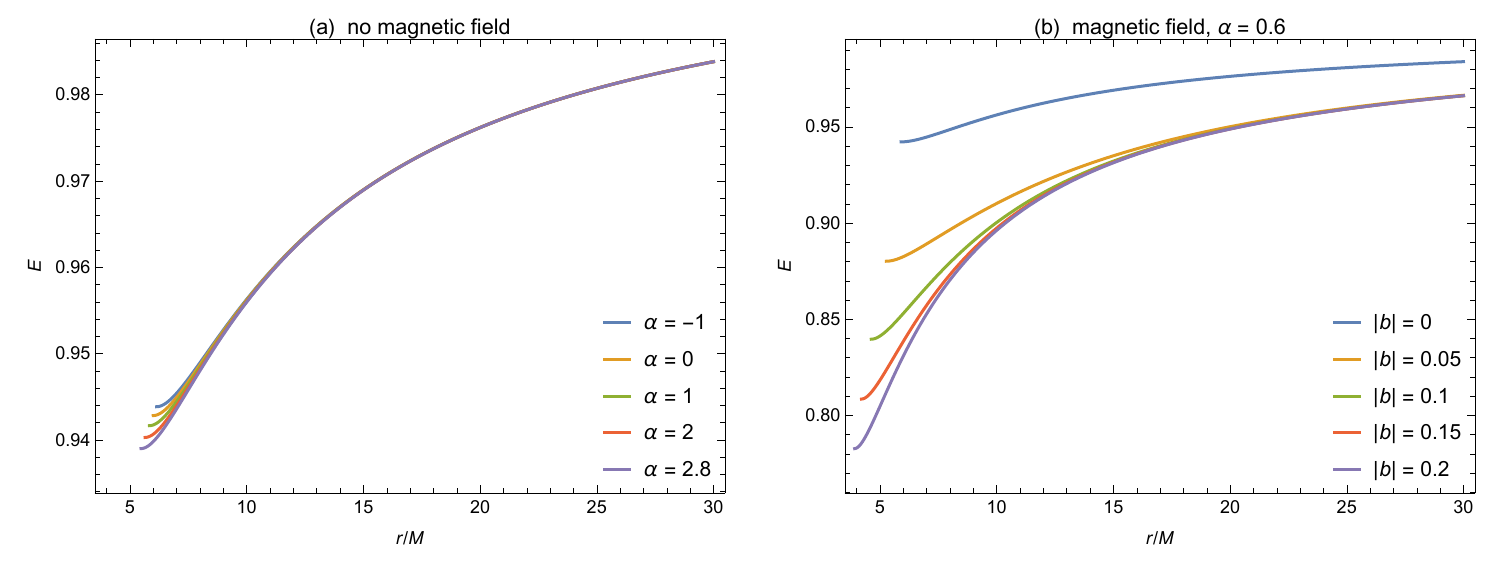}
\caption{Specific energy $E$ of the circular orbits. The panels are as in
Fig.~\ref{fig:omega}. The inner end point of each curve corresponds to the
ISCO and its value gives the efficiency $\eta=1-E_{\rm ISCO}$.}\label{fig:energy}
\end{figure*}

\subsection{Orbital quantities}

The orbital angular velocity is shown in Fig.~\ref{fig:omega}. Without magnetic
field the curves for different values of $\alpha$ are almost indistinguishable.
This is not surprising, since the correction term in $f$ decays as $r^{-4}$ and
at the ISCO it is of the order of $10^{-3}$ even for $\alpha=2.8$. The motion of
the gas in the disk is thus essentially Keplerian for every value of $\alpha$.
On the other hand, the magnetic coupling reduces $\Omega$ considerably at all
radii. The reason is that part of the centripetal force is now provided by the
Lorentz force. For example, for $|b|=0.20$ the angular velocity at $r=10M$ is
only $17\%$ of the value without magnetic field, and at $r=20M$ it is $6\%$. As
we shall see, this effect is responsible for most of the results that follow.

\begin{figure*}[t]\includegraphics[width=0.98\textwidth]{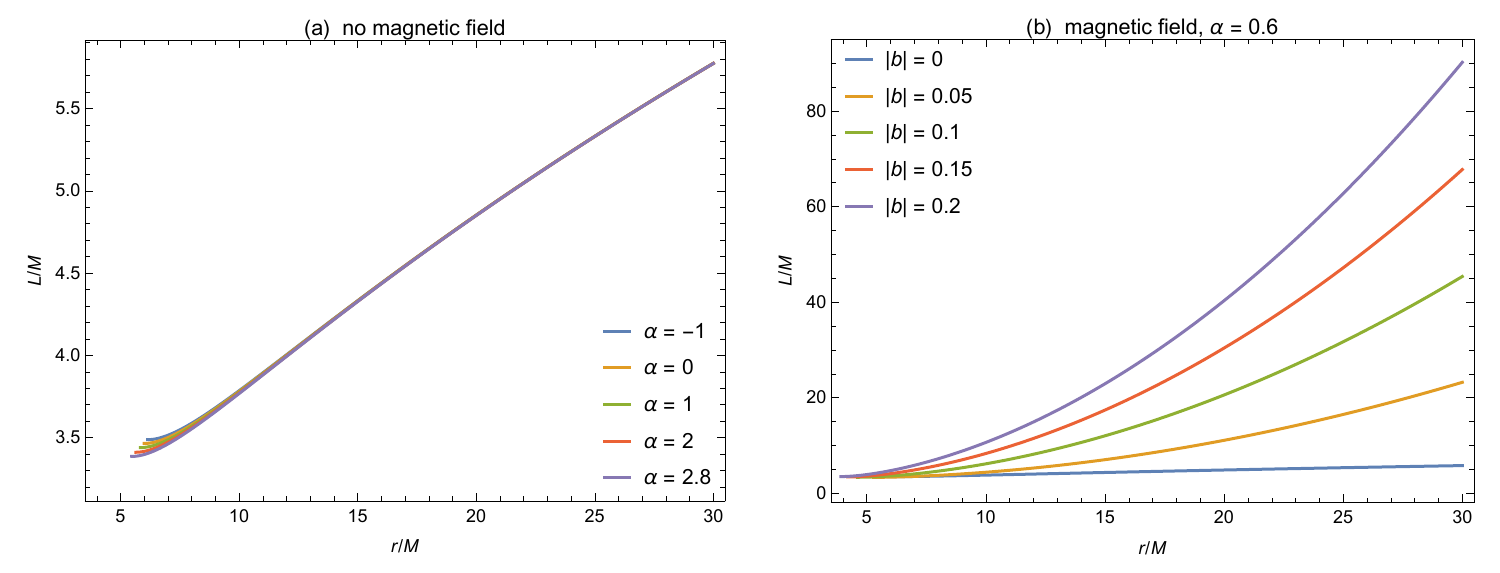}
\caption{Canonical specific angular momentum $L/M$ of the circular orbits, with
panels as in Fig.~\ref{fig:omega}. Notice the different vertical scales in the
two panels. The quadratic growth in the right panel is due to the term
$\tfrac b2 r^2$ in Eq.~\eqref{eq:ELO}.}\label{fig:angmom}
\end{figure*}

The specific energy is shown in Fig.~\ref{fig:energy}. In all cases it grows
monotonically from $E_{\rm ISCO}$ to unity. Over the whole range of $\alpha$ the
value of $E_{\rm ISCO}$ changes only from $0.9438$ to $0.9390$, while for
$|b|=0.20$ it drops to $0.7827$. Magnetized orbits are therefore much more
tightly bound, which, as we shall see, implies a larger efficiency of the disk.
At the same time the ISCO moves from $6.145M$ to $5.477M$ when $\alpha$ goes
from $-1$ to $2.8$, and down to $3.919M$ for $|b|=0.20$.

The angular momentum is shown in Fig.~\ref{fig:angmom}. Again, without magnetic
field the curves nearly coincide. With the magnetic field $L$ grows
quadratically at large radii and reaches $L\simeq90M$ at $r=30M$ for
$|b|=0.20$, while the kinetic part $X$ decays as $M/(br)$. In other words, most
of the angular momentum of the disk is stored in the magnetic term rather than
in the rotation of the gas, and this is why $\Omega$ can be strongly reduced
while the orbits remain stable.

\section{Disk observables}\label{sec:disk}

\begin{figure*}[t]\includegraphics[width=0.98\textwidth]{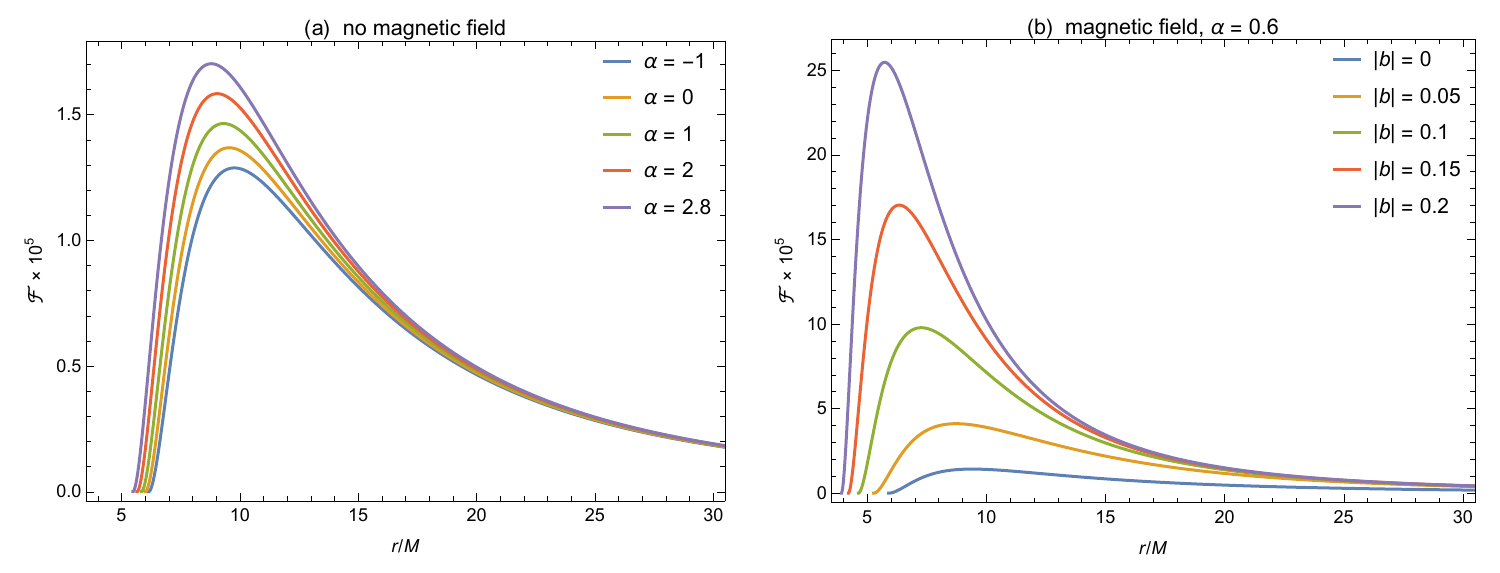}
\caption{Radiative flux $\mathcal{F}(r)$ of the Novikov--Thorne disk in units of
$\dot m/M^2$, with panels as in Fig.~\ref{fig:omega}. The flux vanishes at the
ISCO, reaches a maximum a few $M$ outside it and then decays as $r^{-3}$.
Notice the different vertical scales.}\label{fig:flux}
\end{figure*}

\subsection{Radiative flux and temperature}

In the NT model the gas moves on nearly circular orbits and the heat generated
by viscosity is radiated locally as a black body. From the conservation of rest
mass, energy and angular momentum the time averaged radiative flux
is~\cite{PageThorne1974,Nieto2025}
\begin{equation}
\mathcal{F}(r)=-\frac{\dot m}{4\pi\sqrt{-g}}
\frac{\Omega_{,r}}{\left(E-\Omega L\right)^2}
\int_{r_{\rm ISCO}}^{r}\left(E-\Omega L\right)L_{,r}\,dr' ,
\label{eq:flux}
\end{equation}
where $\dot m$ is the accretion rate, which we take to be constant, and
$\sqrt{-g}=r$ is the determinant of the metric restricted to the equatorial
$(t,r,\phi)$ hypersurface. Here $E$, $L$ and $\Omega$ are the quantities of the
charged orbits given by Eq.~\eqref{eq:ELO}. It is important to notice that
Eq.~\eqref{eq:flux} still holds in the present stationary and axially symmetric
test field approximation, provided that $E$ and $L$ are the canonical
quantities conserved by the two Killing symmetries. In fact, since $A_t=0$ the
Lorentz force does no work with respect to the Killing energy, while the axial
symmetry conserves the canonical angular momentum, which includes the term
$\bar qA_\phi$. We also assume that the external field does not carry a net
radial flux of energy or angular momentum and we neglect its backreaction. For
the numerical integration it is useful to notice that
$E-\Omega L=f\left(1-\tfrac{b}{2}X\right)/E$, which is positive over the whole
disk. For $\alpha=b=0$ we recover the standard NT profile, with
$\mathcal{F}_{\max}=1.368\times10^{-5}\dot m/M^2$ at $r=9.55M$.

As shown in Fig.~\ref{fig:flux}, the maximum of the flux increases with $\alpha$,
by about $32\%$ between $\alpha=-1$ and $\alpha=2.8$, and it moves inwards from
$9.77M$ to $8.79M$. The effect of the magnetic coupling is much larger. For
$|b|=0.20$ the maximum is $25.46\times10^{-5}\dot m/M^2$, i.e. almost $18$ times
larger than without the field, and it is located at $5.73M$. Also, the profiles
of the magnetized disks are narrower, since the emission is concentrated in the
inner region.

\begin{figure*}[t]\includegraphics[width=0.98\textwidth]{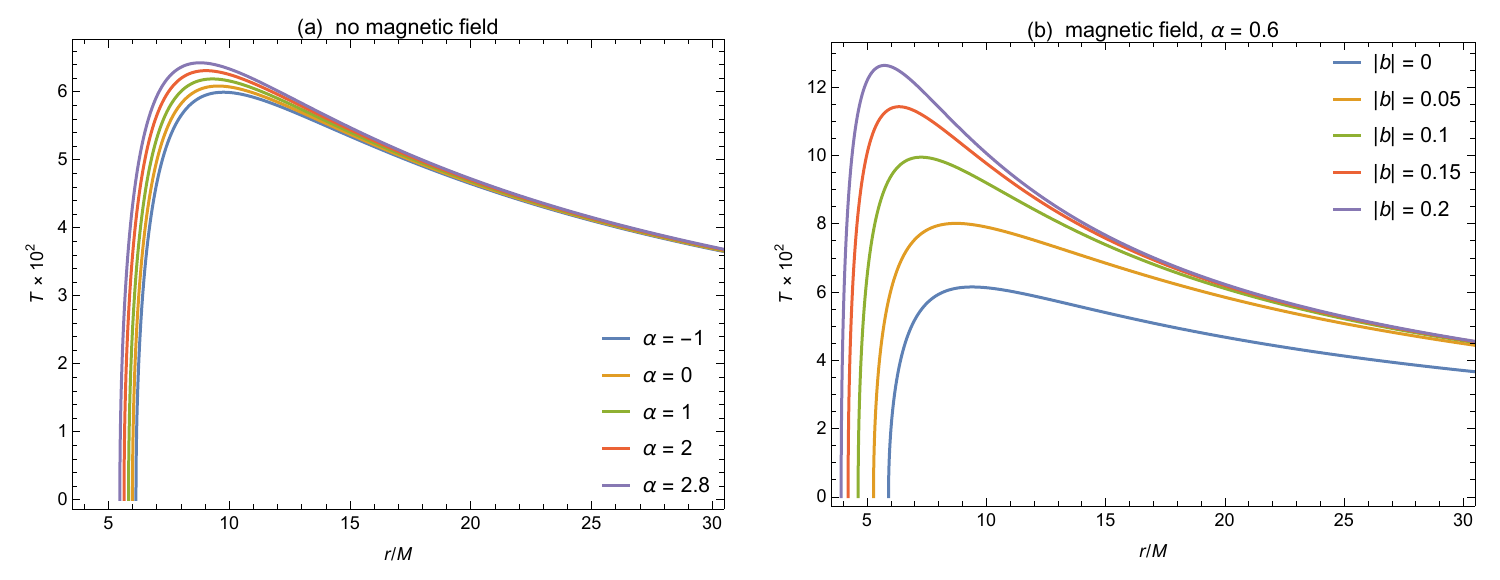}
\caption{Effective temperature $T(r)=(\mathcal{F}/\sigma)^{1/4}$ of the disk, with
panels as in Fig.~\ref{fig:omega}.}
\label{fig:temp}
\end{figure*}

Assuming local thermodynamical equilibrium, the effective temperature of the
disk is
\begin{equation}
T(r)=[\mathcal{F}(r)/\sigma]^{1/4},
\end{equation}
where $\sigma$ is the Stefan--Boltzmann constant. The temperature profiles are
shown in Fig.~\ref{fig:temp}. Of course they follow the same behavior as the
flux, but the differences are reduced by the fourth root. The $32\%$ increase of
the maximum flux with $\alpha$ corresponds to a $7.2\%$ increase of $T_{\max}$,
while the magnetic coupling increases $T_{\max}$ by a factor of about $2$.

\subsection{Redshift factor}

Photons emitted by the disk are redshifted by the gravitational field and by the
motion of the gas before reaching a distant observer. For an emitter on a
circular orbit the time component of the four-velocity is
\begin{equation}
u^t(r)=\frac{1}{\sqrt{f(r)-\Omega^2(r)h(r)}}=\frac{E(r)}{f(r)},
\label{eq:ut}
\end{equation}
where the second equality holds also for charged orbits since $A_t=0$. Then the
redshift factor of a photon with specific angular momentum $\lambda$ about the
symmetry axis, as measured by an observer at infinity, is
\begin{equation}
\tilde g=\frac{E_{\rm obs}}{E_{\rm em}}=\frac{1}{u^t(r)\left[1-\Omega(r)\lambda\right]} .
\label{eq:redshift}
\end{equation}
For an observer looking at the disk face on we have $\lambda=0$ and the redshift
factor reduces to
\begin{equation}
\tilde g=\frac{1}{u^t(r)}=\sqrt{f(r)-\Omega^2(r)h(r)},
\end{equation}
which depends only on the emission radius.

\begin{figure*}[t]\includegraphics[width=0.98\textwidth]{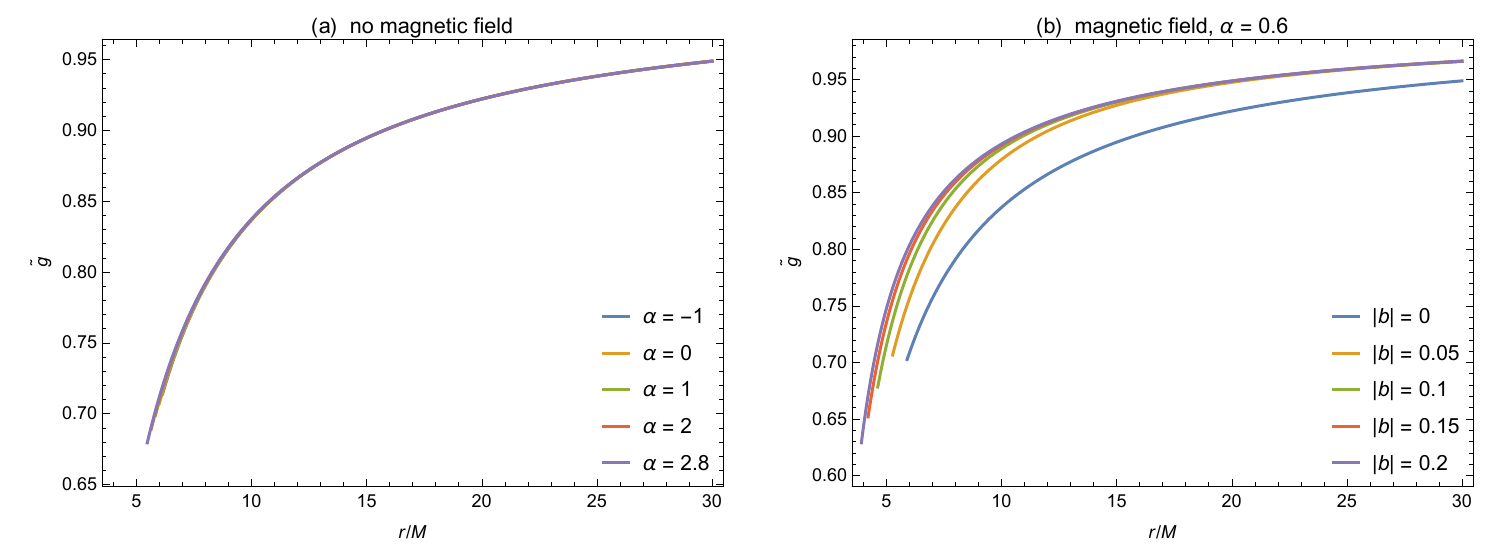}
\caption{Redshift factor $\tilde g=1/u^t$ for photons emitted by the disk and
observed face on, with panels as in Fig.~\ref{fig:omega}. In the Schwarzschild
case $\tilde g=1/\sqrt2$ at the ISCO.}\label{fig:redshift}
\end{figure*}

The face-on redshift factor is shown in Fig.~\ref{fig:redshift}. It goes to
unity at large distances and decreases towards the inner edge of the disk. Both
parameters move the inner edge inwards and therefore the photons emitted from
the innermost part of the disk are more redshifted, with
$\tilde g_{\rm in}=0.6794$ for $\alpha=2.8$ and $\tilde g_{\rm in}=0.6288$ for
$|b|=0.20$. Notice however that, at a given radius, the magnetized curves lie
above the field free one. This is due to the smaller orbital velocity of the
magnetized disk, which reduces the transverse Doppler effect.

\subsection{Differential and spectral luminosity}

The differential luminosity measured at infinity is~\cite{Nieto2025,Zare2025}
\begin{equation}
\frac{d\mathcal{L}_\infty}{d\ln r}=4\pi r\sqrt{-g}\,E(r)\,\mathcal{F}(r)
=4\pi r^2 E(r)\,\mathcal{F}(r).
\label{eq:difflum}
\end{equation}

\begin{figure*}[t]\includegraphics[width=0.98\textwidth]{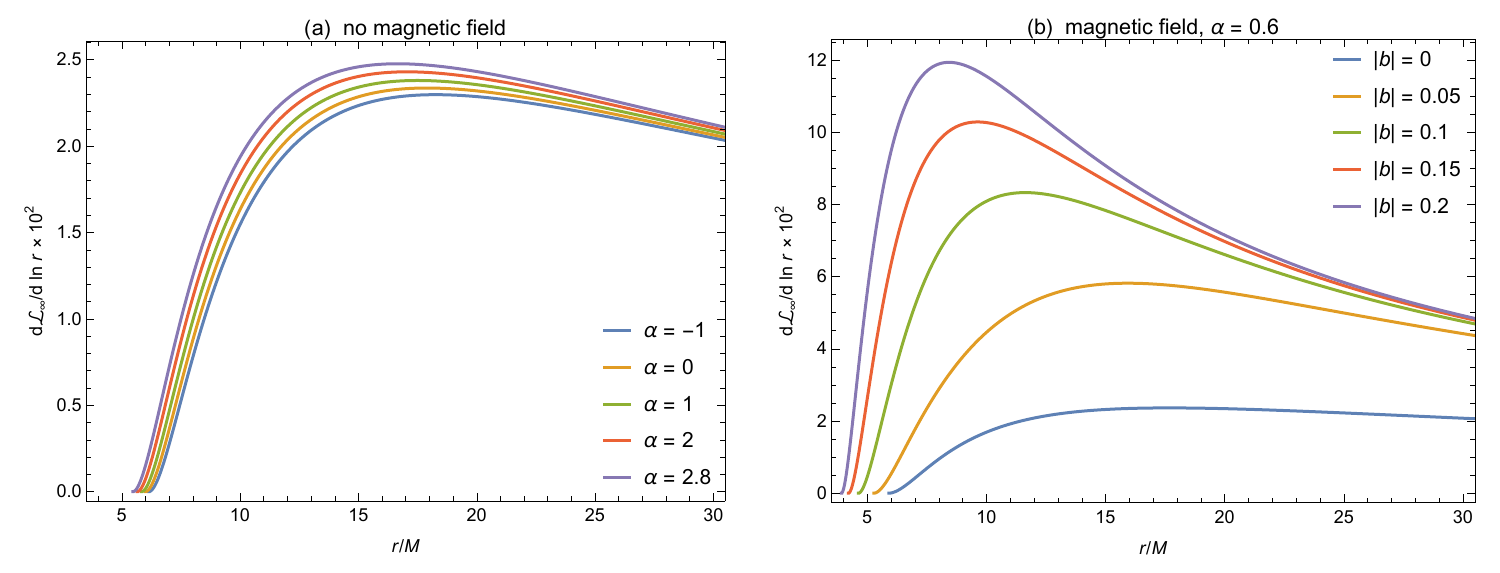}
\caption{Differential luminosity $d\mathcal{L}_\infty/d\ln r$ measured at
infinity, with panels as in Fig.~\ref{fig:omega}. The area under each curve is
the bolometric luminosity, which must be equal to $\eta\dot
m$.}\label{fig:difflum}
\end{figure*}

The differential luminosity is shown in Fig.~\ref{fig:difflum}. Because of the
factor $r^2$ the maximum is located further out with respect to the maximum of
the flux. Integrating over the disk we find that the bolometric luminosity
increases by a factor of $3.75$ between $|b|=0$ and $|b|=0.20$ at $\alpha=0.6$,
while it increases only by $8.6\%$ over the whole range of $\alpha$. As we
discuss below, these ratios must coincide with the corresponding ratios of the
efficiency and in fact they do, within the numerical accuracy. Also, the maximum
of the differential luminosity moves inwards when $|b|$ increases, meaning that
a magnetized disk emits a larger fraction of its luminosity from the region
where relativistic effects are strong.

Assuming that each ring of the disk emits as a black body, the spectral
luminosity observed at infinity is~\cite{Boshkayev2020}
\begin{align}
\nu\mathcal{L}_{\nu,\infty}={}&\frac{15}{\pi^{4}}
\int_{r_{\rm ISCO}}^{\infty}
\left(\frac{d\mathcal{L}_{\infty}}{d\ln r}\right)
\frac{\left[u^t(r)y\right]^{4}}{M^{2}\mathcal{F}(r)}\nonumber\\
&\times\frac{d\ln r}
{\exp\!\left(\dfrac{u^t(r)y}
{\left[M^{2}\mathcal{F}(r)\right]^{1/4}}\right)-1},
\label{eq:spectrum}
\end{align}
where $y=h\nu/kT_*$, $h$ is Planck's constant, $\nu$ is the frequency, $k$ is
Boltzmann's constant and $T_*=[\dot m/(4\pi M^2\sigma)]^{1/4}$ is the
characteristic temperature of the disk.

\begin{figure*}[t]\includegraphics[width=0.98\textwidth]{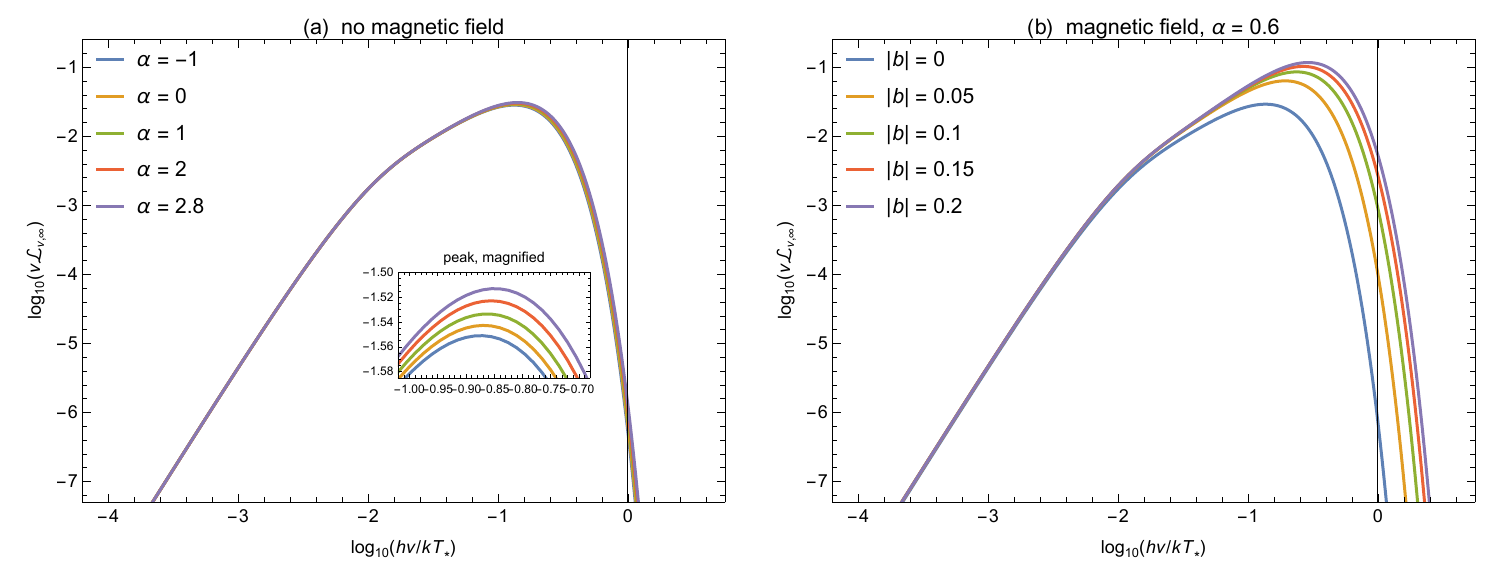}
\caption{Spectral luminosity $\nu\mathcal{L}_{\nu,\infty}$ observed at infinity,
integrated up to $r_{\rm out}=10^3M$, with panels as in Fig.~\ref{fig:omega}.
The inset shows the peak of the spectra in the left panel, where the five
curves differ by only $0.038$~dex (about $9\%$) in height.}
\label{fig:spectrum}
\end{figure*}

The spectral luminosity is shown in Fig.~\ref{fig:spectrum}. The magnetic
coupling increases the maximum of $\nu\mathcal{L}_{\nu,\infty}$ by a factor of
$4$ and moves it to higher frequencies by a factor of about $2$, from
$\log_{10}y=-0.865$ to $-0.537$. Therefore the spectrum of a magnetized disk is
both brighter and harder. On the other hand, the quantum parameter increases
the maximum only by $9\%$ and moves it by $0.025$~dex, so that the inset is
needed to see the differences. In principle the two parameters could then be
distinguished, with $\alpha$ affecting mostly the amplitude of the spectrum and
$b$ its shape. It is interesting to notice that the effect is opposite to the
one found for dark matter halos~\cite{Nieto2025,Zare2025}, where the halo makes
the spectrum redder and less intense. Of course in practice a change in
amplitude of a few percent is degenerate with the accretion rate and the
distance of the source, and therefore observations of the continuum can
constrain $b$ much better than $\alpha$.

\subsection{Radiative efficiency}

The efficiency of the disk in converting accreted mass into radiation
is~\cite{Kurmanov2022}
\begin{equation}
\eta=1-E_{\rm ISCO}.
\label{eq:eta}
\end{equation}
Notice that Eqs.~\eqref{eq:flux}, \eqref{eq:difflum} and~\eqref{eq:eta} are not
independent, since the integral of the differential luminosity over the whole
disk must give $\mathcal{L}_\infty=\eta\,\dot m$. We used this relation as a
check of the numerics. Since at large radii
$\mathcal{F}\simeq3\dot mM/(8\pi r^3)$, truncating the integral at
$r_{\rm out}$ neglects a contribution of order $M/r_{\rm out}$. In fact for
$\alpha=b=0$ we find $\mathcal{L}_\infty/(\eta\dot m)=0.9759$, $0.9987$ and
$0.99974$ for $r_{\rm out}=10^3M$, $2\times10^4M$ and $10^5M$ respectively, and
the same happens for all the other cases.

\begin{figure*}[t]\includegraphics[width=0.98\textwidth]{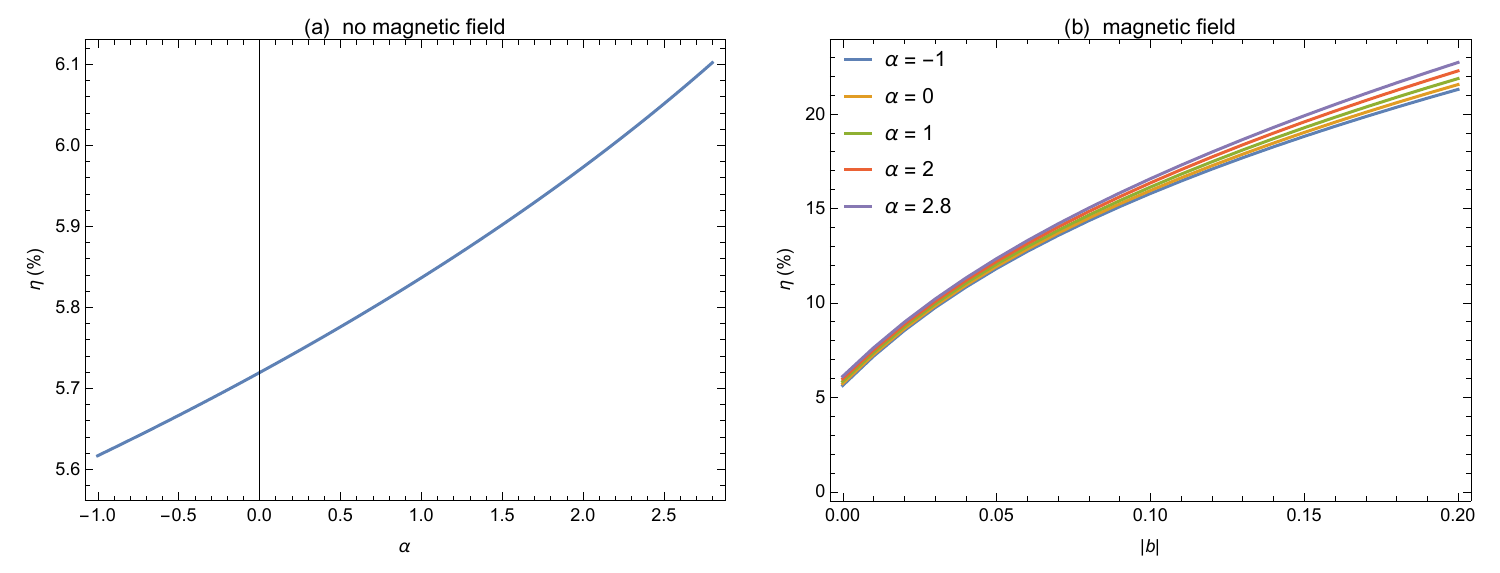}
\caption{Efficiency $\eta=1-E_{\rm ISCO}$ of the disk. Left: as a function of
$\alpha$ for $|b|=0$. Right: as a function of the magnetic coupling
$|b|=|\bar qB|M$ for different values of $\alpha$. Notice the different vertical
scales.}
\label{fig:eta}
\end{figure*}

The efficiency is shown in Fig.~\ref{fig:eta}. Over the whole range of $\alpha$
it changes only from $5.617\%$ to $6.102\%$, i.e. by less than half a
percentage point, while the magnetic coupling alone increases it from $5.787\%$
to $21.728\%$ for $|b|=0.20$. The curves for different values of $\alpha$ in the
right panel remain ordered and almost parallel, and their separation never
exceeds $1.43$ percentage points. We also checked that $\eta$ increases
monotonically with both parameters.

\section{Iron K$\alpha$ line}\label{sec:iron}

\begin{figure*}[t]\includegraphics[width=0.98\textwidth]{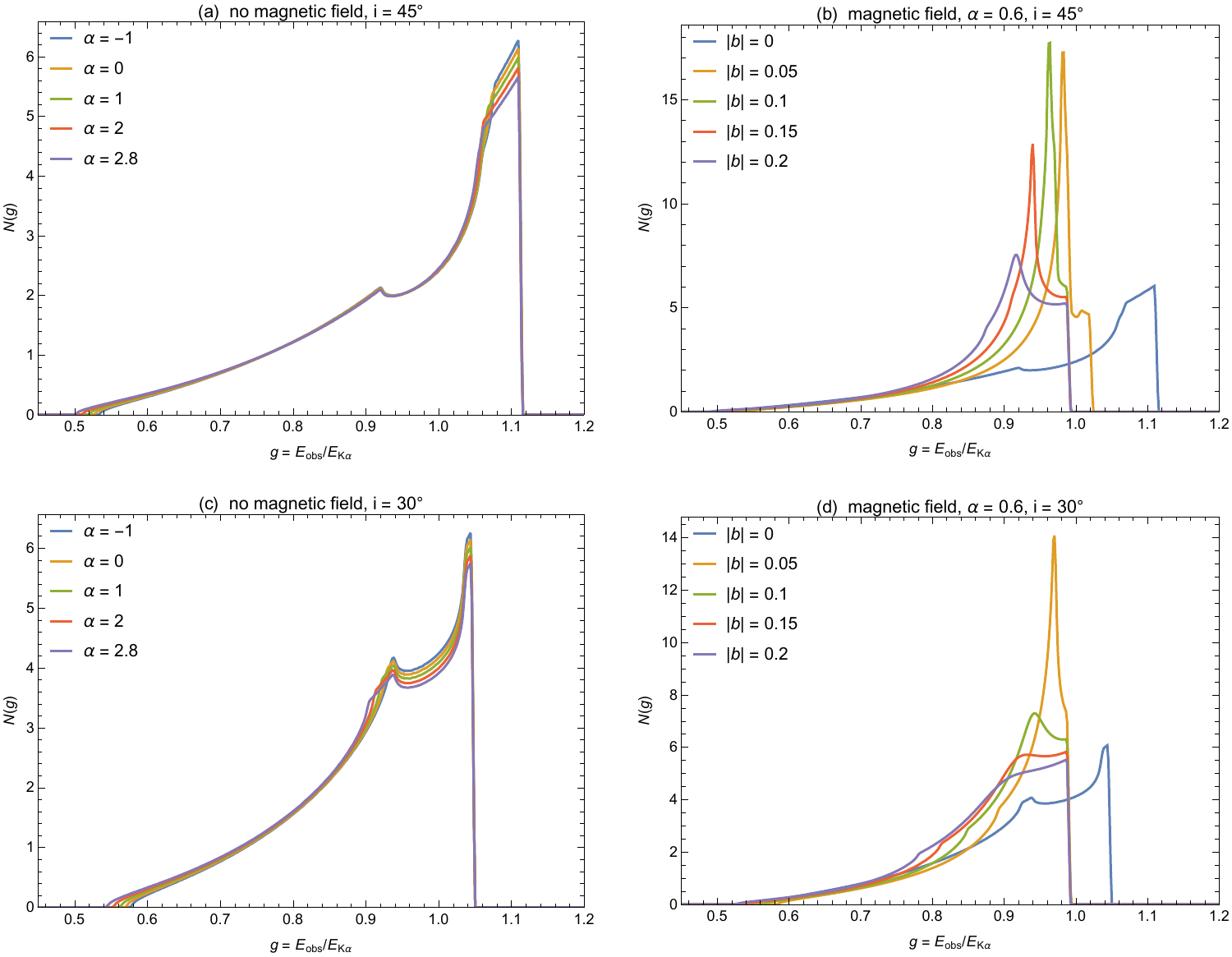}
\caption{Iron K$\alpha$ line profiles for an observer with inclination
$i=45^\circ$ (top) and $i=30^\circ$ (bottom), for emissivity index $q=3$ and a
disk extending from $r_{\rm ISCO}$ to $r_{\rm ISCO}+100M$. The panels are as in
Fig.~\ref{fig:omega}. The photon flux $N(g)$ is normalized to unit area, with
$g=E_{\rm obs}/E_{K\alpha}$.}\label{fig:iron}
\end{figure*}

The thermal spectrum is not the only feature of the X-ray emission produced in
the inner part of the disk. The fluorescent iron K$\alpha$ line is emitted at
$E_{K\alpha}=6.4$~keV in the rest frame of the gas and it is intrinsically
narrow. Therefore the shape observed at infinity carries information on the
Doppler shift of the orbiting gas, on the gravitational redshift and on light
bending~\cite{Fabian1989,Laor1991,Reynolds2014}. Here we follow the treatment
of Ref.~\cite{Urmanov2025}, where the line was computed for the black hole
obtained from dust collapse in asymptotic safety~\cite{BMP2024}. The comparison
between the two models is quite natural. In fact, the mass function of the
asymptotically safe black hole behaves as $M(r)\simeq M-3M^2\xi/r^3$ at large
distances, so that its lapse coincides with Eq.~\eqref{eq:f} at first order if
we set $\alpha=6\xi$.

The photon flux received by a distant observer is~\cite{Bambi2012,Urmanov2025}
\begin{equation}
N(E_{\rm obs})=\frac{1}{E_{\rm obs}}\int I_{\rm obs}\,d\Omega_{\rm obs}
=\frac{1}{E_{\rm obs}}\int \tilde g^3 I_{\rm e}(E_{\rm e})\,d\Omega_{\rm obs},
\label{eq:Nobs}
\end{equation}
where $d\Omega_{\rm obs}$ is the element of solid angle on the observer's sky,
$\tilde g=E_{\rm obs}/E_{\rm e}$ is the redshift factor~\eqref{eq:redshift}, and
we used Liouville's theorem, $I_{\rm obs}=\tilde g^3I_{\rm e}$. As usual we
assume that the emission is monochromatic and isotropic in the rest frame of
the gas, with a power-law radial profile
\begin{equation}
I_{\rm e}\propto\frac{\delta(E_{\rm e}-E_{K\alpha})}{r^{q}} ,
\label{eq:emissivity}
\end{equation}
with $q=3$~\cite{WilkinsFabian2012}, and that the disk extends from
$r_{\rm in}=r_{\rm ISCO}$ to $r_{\rm out}=r_{\rm ISCO}+100M$, as in
Ref.~\cite{Urmanov2025}.

It is important to notice that the magnetic field enters Eq.~\eqref{eq:Nobs}
only through the emitters. Photons are neutral and move along the null
geodesics of the metric~\eqref{eq:metric}, which depend only on $\alpha$. We
trace them backwards from the image plane of an observer at infinity with
inclination $i$, where each point has polar coordinates $(\rho,\vartheta)$ and
$\rho$ is the impact parameter. Each photon moves in a plane and, with
$x=M/r$, its trajectory is given by
\begin{equation}
\frac{d^2x}{d\phi^2}=-x+3x^2-\frac{3\alpha}{M^2}\, x^5 ,
\label{eq:orbit}
\end{equation}
with $x=0$ and $dx/d\phi=M/\rho$ at $\phi=0$. The photon crosses the equatorial
plane for the first time at $\phi_0={\rm arccot}(-\sin\vartheta\tan i)\in(0,\pi)$,
and the corresponding radius is the emission point~\cite{Luminet1979}. The gas, on
the other hand, moves on the charged circular orbits of Sec.~\ref{sec:orbits}.
Then, since $\lambda=-\rho\cos\vartheta\sin i$, from Eqs.~\eqref{eq:ut}
and~\eqref{eq:redshift} we get
\begin{equation}
\tilde g=\frac{f(r)}{E(r)\left[1+\Omega(r)\,\rho\cos\vartheta\sin i\right]} ,
\label{eq:gline}
\end{equation}
where $E$ and $\Omega$ are given by Eq.~\eqref{eq:ELO}. We integrate
Eq.~\eqref{eq:orbit} with a fourth-order Runge--Kutta method for a grid of about
$3\times10^7$ points on the image plane and then construct $N(E_{\rm obs})$ as a
histogram in $\tilde g$. Photons that cross the photon sphere inwards are not
followed further. For black holes they are captured, while for horizonless
configurations they are reflected by the inner potential barrier, but we checked
that in this case they would add only about $1\%$ of the flux. Similarly, the
higher order images, coming from the subsequent crossings of the equatorial
plane at $\phi_0+n\pi$, contribute between $0.5\%$ and $2\%$ of the flux and are
not included in Fig.~\ref{fig:iron}.

As a test of the code we recomputed the line profiles of
Ref.~\cite{Urmanov2025} for the asymptotically safe black hole with $\xi=0.2$,
$0.4$ and $0.6$ (in units of $M^2$), for $i=30^\circ$, $45^\circ$ and $q=3$,
$4$, which were obtained there with the ray tracing code
Gradus.jl~\cite{Gradus}. The two calculations agree within the resolution of the
published figures, including the positions of the red and blue edges and the
heights of the peaks.

The profiles for the qOS black hole are shown in Fig.~\ref{fig:iron}. Without
magnetic field the line has the familiar shape, with an extended red tail due
to the gravitational redshift in the inner part of the disk and a sharp blue
peak due to the gas moving towards the observer. As expected from the behavior
of the ISCO, the quantum parameter affects mostly the red part of the line. When
$\alpha$ goes from $-1$ to $2.8$ the inner edge of the disk moves inwards and
the red edge of the line moves from $g\simeq0.534$ to $0.502$ for $i=45^\circ$
(and from $0.578$ to $0.545$ for $i=30^\circ$), while the blue peak becomes
about $10\%$ lower. The position of the blue edge does not change, since it is
determined by the Doppler shift at larger radii, where the correction term is
negligible. These trends are the same found in Ref.~\cite{Urmanov2025} for the
asymptotically safe black hole, and in fact the profiles obtained for $\alpha$
and for $\xi=\alpha/6$ are almost identical. For example, for $\alpha=2.4$ and
$\xi=0.4$ the two profiles differ by less than $0.5\%$ of the height of the
blue peak, which is more than ten times smaller than their deviation from the
Schwarzschild profile.

The effect of the magnetic field is much stronger. 
As discussed in Sec. Sec.~\ref{sec:orbits}, the orbital velocity of the gas is reduced because the outward Lorentz force partially counteracts the gravitational attraction.
 Then the Doppler
splitting, which is responsible for the double peaked shape of the line,
becomes much smaller. Already for $|b|=0.05$ the blue edge moves from
$g\simeq1.11$ to $1.02$ for $i=45^\circ$, and for larger values it is fixed at
$g\simeq0.99$ by the gravitational redshift at the outer edge of the disk,
$\sqrt{f(r_{\rm out})}$. At the same time the ISCO moves inwards and the
gravitational redshift at the inner edge becomes larger. For $|b|\gtrsim0.1$
this compensates the smaller Doppler shift and the red edge moves to lower
energies, down to $g\simeq0.49$ for $|b|=0.20$ at $i=45^\circ$. As a result, the
blue peak of the field free case is replaced by a single peak below the rest
frame energy, which moves to lower energies and becomes broader as $|b|$
increases, from $g\simeq0.98$ for $|b|=0.05$ to $g\simeq0.92$ for $|b|=0.20$ at
$i=45^\circ$. Correspondingly, the mean energy of the line decreases from
$0.95E_{K\alpha}$ to $0.88E_{K\alpha}$. It is worth noticing that the blue part
of the magnetized profiles resembles the one of a field free disk seen almost
face on. For example, for $\alpha=0.6$, $b=0$ and $i=10^\circ$ we find the blue
edge at $g\simeq1.00$ and the peak at $g\simeq0.97$. However, the red tail of the
magnetized line is much longer, with the red edge at $g\simeq0.49$--$0.55$ for
$i=45^\circ$ against $g\simeq0.65$ for the face-on disk, so that the two cases
can in principle be distinguished. For $q=4$ the emission is more concentrated
in the inner part of the disk and we find the same behavior, with a larger
effect on the red tail.

Therefore the iron line responds to the two parameters in a different way with
respect to the thermal spectrum. The quantum parameter only shifts the red edge
of the line by a few percent, which is comparable with the effect of
$\xi$ in the asymptotically safe case and is likely to be degenerate with the
emissivity profile and with the spin of the black hole~\cite{Urmanov2025}. On
the other hand, the suppression of the blue peak is a clear signature of the
magnetic coupling, since it can not be produced by a change of the geometry
alone at moderate inclinations.

\section{Gravitational capture cross sections}\label{sec:capture}

The capture of test particles by a black hole is one of the classical problems
of black hole physics~\cite{Misner:1973prb,Chandrasekhar:1985kt}. For
a Schwarzschild black hole the capture cross section for photons is
$\sigma=27\pi M^2$, while for massive particles that are slowly moving at
infinity it grows as $\sigma\simeq16\pi M^2/v^2$. Both values are much larger
than the geometric cross section of the horizon, $4\pi M^2$, as a consequence of
the focusing effect of gravity~\cite{Shapiro1983,FrolovNovikov1998}. The capture
cross section determines the rate at which a black hole absorbs matter from its
environment, such as gas, stars or dark matter particles~\cite{Sadeghian:2013laa},
and it is closely related to other observables. In fact, the photon capture
cross section is fixed by the critical impact parameter, which also defines the
size of the shadow seen by a distant observer~\cite{Synge:1966okc,Luminet1979,EHT2019},
while at high frequencies the absorption cross section of waves oscillates
around the same geometric value~\cite{Sanchez:1977si,Decanini:2011xi}. It is important
to notice that the capture is controlled by the maximum of the effective
potential rather than by the ISCO. Therefore the cross sections probe the
geometry closer to the center than the disk, and we may expect the quantum
corrections to be more visible here. In the following we compute the capture
cross sections for massless, massive and charged particles, following the
approach of Refs.~\cite{Zakharov1994,Sucu2025}.

The capture cross section is defined as
\begin{equation}
\sigma_{\rm capt}=\pi\,\beta_{\max}^2 ,
\label{eq:sigmadef}
\end{equation}
where $\beta_{\max}$ is the largest impact parameter for which the particle is
captured.

\subsection{The marginally bound orbit}\label{sec:mb}

\begin{figure*}[t]\includegraphics[width=0.98\textwidth]{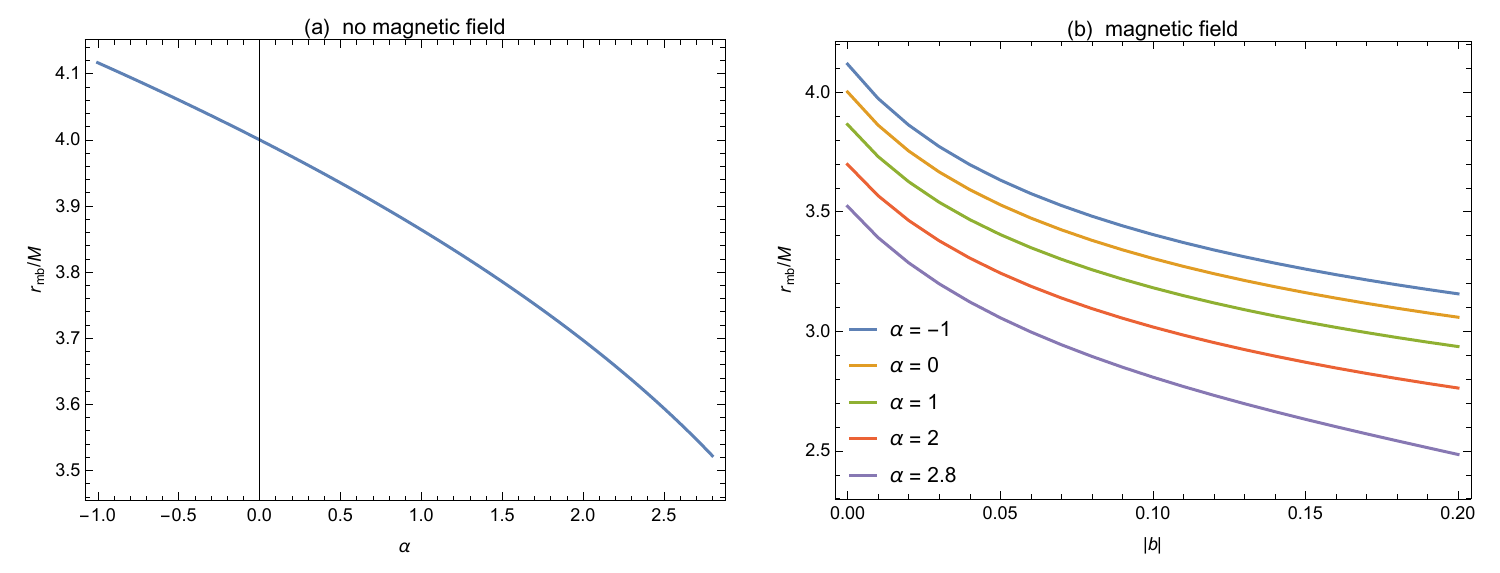}
\caption{Radius of the marginally bound orbit, obtained from
Eqs.~\eqref{eq:mbgeneral} and~\eqref{eq:mbpoly}. Left: no magnetic field.
Right: as a function of the magnetic coupling for different values of
$\alpha$.}\label{fig:rmb}
\end{figure*}

For particles that are slowly moving at infinity the threshold for capture is
given by the marginally bound orbit, i.e. the circular orbit with $E=1$. Its
angular momentum determines the maximum impact parameter as
$\beta_{\max}\simeq L(r_{\rm mb})/(m v_\infty)$. Setting $E=1$ in
Eq.~\eqref{eq:ELO} we get $X^2=r^2(1-f)/f$, and using Eq.~\eqref{eq:quad} we
obtain the condition
\begin{equation}
\Big[\left(f'r-2f\right)(1-f)+f f' r\Big]^2=4f^3b^2r^2(1-f).
\label{eq:mbgeneral}
\end{equation}
For $b=0$ the right hand side vanishes and Eq.~\eqref{eq:mbgeneral} reduces to
$2f^2-2f+rf'=0$, which gives the polynomial
\begin{equation}
r^7-4Mr^6+\alpha Mr^4+4\alpha M^2r^3-\alpha^2M^3=0 .
\label{eq:mbpoly}
\end{equation}
This has only one root outside the photon sphere and for $\alpha=b=0$ we recover
$r_{\rm mb}=L_{\rm mb}=4M$. In the non-relativistic limit the capture cross
section, normalized to the Schwarzschild value, is then
\begin{equation}
\frac{v^2\sigma_{\rm capt}}{16\pi M^2}=\frac{L_{\rm mb}^2}{16M^2}.
\label{eq:signr}
\end{equation}

The radius of the marginally bound orbit is shown in Fig.~\ref{fig:rmb}. Without
magnetic field it decreases from $4.117M$ for $\alpha=-1$ to $3.522M$ for
$\alpha=2.8$. This is a $14.4\%$ reduction, larger than the $10.9\%$ reduction
of the ISCO, since $r_{\rm mb}$ is closer to the center where the correction
term is larger. The magnetic field decreases it further, to $2.989M$ for
$|b|=0.20$. Notice that the curves in the right panel are well separated, so
that, contrary to the efficiency, $r_{\rm mb}$ is not nearly degenerate in
$\alpha$. The marginally bound orbit ceases to exist when it reaches the photon
sphere. Setting $r=r_{\rm ph}$ in Eq.~\eqref{eq:mbgeneral} and using $f'r=2f$ we
obtain
\begin{equation}
b_{\max}^{2}=\frac{2r_{\rm ph}-3M}{r_{\rm ph}^{2}\left(r_{\rm ph}+3M\right)} ,
\label{eq:bmax}
\end{equation}
which gives $b_{\max}^2=1/18$ for Schwarzschild and goes from $|b|=0.233$ for
$\alpha=-1$ to $|b|=0.240$ for $\alpha=2.8$, i.e. just outside the range
considered here.

\subsection{Massless particles}

For null geodesics the photon sphere is located at the extremum of $f/h$,
\begin{equation}
\left.\frac{d}{dr}\frac{f(r)}{h(r)}\right|_{r_{\rm ph}}=0
\quad\Longleftrightarrow\quad
r^4-3Mr^3+3\alpha M^2=0 ,
\label{eq:photonsphere}
\end{equation}
and $r_{\rm ph}$ is the largest positive root. As mentioned before, the double
root of this quartic determines $\alpha_{\max}$ in Eq.~\eqref{eq:alphamax}.
Since photons are neutral the Wald field plays no role and
Eq.~\eqref{eq:photonsphere} depends only on $\alpha$. The critical impact
parameter and the cross section are
\begin{equation}
b_{\rm cr}=\frac{r_{\rm ph}}{\sqrt{f(r_{\rm ph})}},\qquad
\sigma_{\rm ph}=\pi b_{\rm cr}^2 ,
\label{eq:sigmaph}
\end{equation}
where $b_{\rm cr}$ is also the radius of the shadow. For $\alpha=0$ we have
$r_{\rm ph}=3M$, $b_{\rm cr}=3\sqrt3M$ and $\sigma_{\rm ph}=27\pi M^2$, which we
use to normalize the results (see Table~\ref{tab:capture}).

From Eq.~\eqref{eq:photonsphere} we can write $\alpha$ in terms of $r_{\rm ph}$
as $\alpha=r_{\rm ph}^3(3M-r_{\rm ph})/(3M^2)$. Substituting this into the lapse
we get $f(r_{\rm ph})=2/3-M/r_{\rm ph}$, so that the cross section takes the
simple form
\begin{equation}
\sigma_{\rm ph}=\frac{3\pi r_{\rm ph}^3}{2r_{\rm ph}-3M} .
\label{eq:sigmaphclosed}
\end{equation}
This gives $27\pi M^2$ for $r_{\rm ph}=3M$ and increases monotonically with
$r_{\rm ph}$ for $r_{\rm ph}>9M/4$. Therefore the decrease of the photon sphere
with $\alpha$ implies directly a smaller shadow.

\begin{figure*}[t]\includegraphics[width=0.98\textwidth]{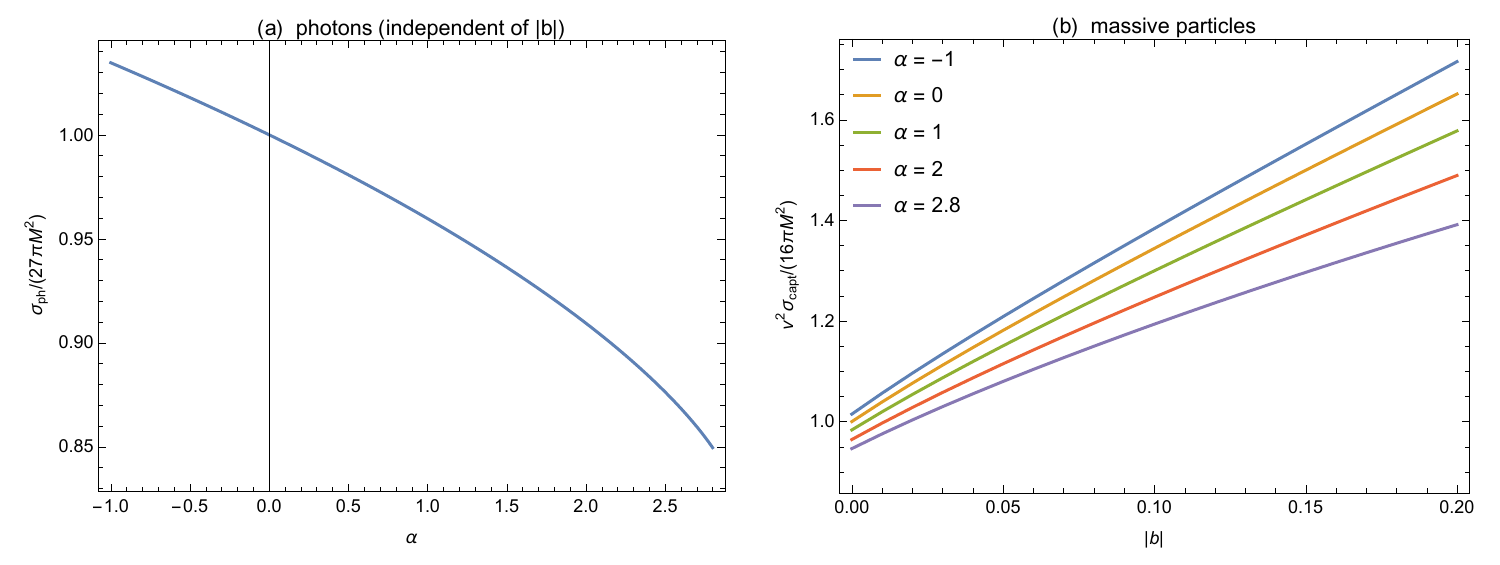}
\caption{Capture cross sections. Left: massless particles, normalized to the
Schwarzschild value $27\pi M^2$. This does not depend on $|b|$. Right: massive
particles in the non-relativistic limit~\eqref{eq:signr} as a function of the
magnetic coupling, for different values of $\alpha$.}\label{fig:capture}
\end{figure*}

It is worth noticing that the equations for the ISCO, the marginally bound orbit
and the photon sphere, i.e. Eqs.~\eqref{eq:iscopoly}, \eqref{eq:mbpoly}
and~\eqref{eq:photonsphere}, are of degree seven, seven and four in $r$, while
they are only quadratic, quadratic and linear in $\alpha$. Then they can be
inverted exactly as
\begin{align}
\alpha_{\rm ISCO}(r)&=\frac{r^3}{24M^2}
  \Big(4r+9M\nonumber\\
  &\qquad-\sqrt{16r^2+120Mr-207M^2}\,\Big),\label{eq:invisco}\\
\alpha_{\rm mb}(r)&=\frac{r^3}{2M^2}
  \left(r+4M-\sqrt{r\left(r+12M\right)}\right),\label{eq:invmb}\\
\alpha_{\rm ph}(r)&=\frac{r^3\left(3M-r\right)}{3M^2},\label{eq:invph}
\end{align}
where the signs are fixed by requiring that $\alpha=0$ gives $6M$, $4M$ and
$3M$ respectively. These relations give the left panels of Figs.~\ref{fig:rmb}
and~\ref{fig:capture} in parametric form, and the maximum of
Eq.~\eqref{eq:invisco} gives the value of $\alpha^{\max}_{\rm ISCO}$ quoted in
Sec.~\ref{sec:bg}. Similarly, Eq.~\eqref{eq:mbgeneral} is linear in $b^2$ and
can be inverted as
\begin{equation}
b^2=\frac{\Big[\left(f'r-2f\right)(1-f)+ff'r\Big]^2}{4f^3r^2(1-f)} .
\label{eq:invb}
\end{equation}
For small $\alpha$ the radii can also be obtained as power series in $\alpha$.
These expansions, together with the corresponding expansions of the critical
impact parameter and of the cross sections, are given in
Appendix~\ref{app:series}. For example, at first order the photon capture cross
section is simply $\sigma_{\rm ph}\simeq\pi\left(27M^2-\alpha\right)$.

\begin{table}[b]
\caption{Photon sphere, critical impact parameter and capture cross sections.
The photon quantities do not depend on the magnetic coupling. The last column is
the non-relativistic cross section~\eqref{eq:signr} for $|b|=0$. The cross
sections are normalized to their Schwarzschild values.}
\label{tab:capture}
\begin{ruledtabular}
\begin{tabular}{ccccc}
$\alpha$ & $r_{\rm ph}/M$ & $b_{\rm cr}/M$
 & $\sigma_{\rm ph}/(27\pi M^2)$
 & $v^2\sigma_{\rm capt}/(16\pi M^2)$\\
\hline
$-1.0$ & 3.1006 & 5.2854 & 1.0346 & 1.0148\\
 0.0   & 3.0000 & 5.1962 & 1.0000 & 1.0000\\
 1.0   & 2.8736 & 5.0904 & 0.9597 & 0.9834\\
 2.0   & 2.6927 & 4.9552 & 0.9094 & 0.9643\\
 2.8   & 2.3649 & 4.7894 & 0.8496 & 0.9463\\
\end{tabular}
\end{ruledtabular}
\end{table}

The capture cross sections are shown in Fig.~\ref{fig:capture}. Here the
quantum correction has the opposite effect with respect to the disk. The photon
sphere decreases from $3.101M$ to $2.365M$ and the critical impact parameter from
$5.285M$ to $4.789M$, so that $\sigma_{\rm ph}$ decreases by $18\%$, from $1.035$
to $0.850$ in units of $27\pi M^2$. Since photons are neutral and the test
field does not change the geometry, $\sigma_{\rm ph}$ does not depend on $|b|$.
Then a measurement of the shadow would constrain $\alpha$ alone, independently
of the magnetic coupling which dominates the disk observables. The cross
section for massive particles, instead, increases by $62\%$ between $|b|=0$ and
$|b|=0.20$, while it decreases by $6.8\%$ over the range of $\alpha$.

\subsection{Massive particles at arbitrary velocity}

A neutral massive particle coming from infinity with velocity $v$ has
$E=\gamma=(1-v^2)^{-1/2}$ and impact parameter $\beta=L/(\gamma v)$. The
particle is captured if
\begin{equation}
\gamma^2>\max_{r>r_h}\,f(r)\left(1+\frac{L^2}{r^2}\right),
\label{eq:capcond}
\end{equation}
which defines a critical angular momentum $L_{\rm cr}(\alpha,v)$ and the cross
section
\begin{equation}
\sigma_{\rm cap}(\alpha,v)=\pi\,\frac{L_{\rm cr}^2(\alpha,v)}{\gamma^2v^2}.
\label{eq:sigmamassive}
\end{equation}
For $v\to1$ we recover the photon cross section and for $v\to0$ the result of
Eq.~\eqref{eq:signr}, and both limits are reproduced numerically to better than
$10^{-4}$.

In fact the relation between $v$ and $\sigma_{\rm cap}$ can be written in
parametric form. When the condition~\eqref{eq:capcond} is saturated we have
$V^2=\gamma^2$ and $\left(V^2\right)'=0$ at the same radius, which means that
the critical trajectory approaches the unstable circular orbit with energy
$\gamma$. Then, using the relations for circular geodesics,
$L^2=f'r^3/(2f-f'r)$ and $E^2=2f^2/(2f-f'r)$, and denoting by $r_c$ the radius
of that orbit, we get
\begin{equation}
v^2=\frac{2f^2-2f+rf'}{2f^2}\bigg|_{r_c},\qquad
\frac{\sigma_{\rm cap}}{\pi}=\frac{r^3f'}{2f^2-2f+rf'}\bigg|_{r_c},
\label{eq:sigmaparam}
\end{equation}
so that
\begin{equation}
\frac{v^2\sigma_{\rm cap}}{\pi}=\frac{r^3f'}{2f^2}\bigg|_{r_c}.
\label{eq:vsigmaparam}
\end{equation}
When $r_c$ goes from $r_{\rm mb}$ to $r_{\rm ph}$ these relations give the whole
curve. For $r_c=r_{\rm mb}$ the denominator in Eq.~\eqref{eq:sigmaparam}
vanishes because of Eq.~\eqref{eq:mbpoly}, so that $v=0$ and we recover
Eq.~\eqref{eq:signr}. For $r_c=r_{\rm ph}$ we have $f'r=2f$, which gives $v=1$
and $\sigma_{\rm cap}=\pi r_{\rm ph}^2/f(r_{\rm ph})$, i.e.
Eq.~\eqref{eq:sigmaph}.

\begin{figure*}[t]\includegraphics[width=0.98\textwidth]{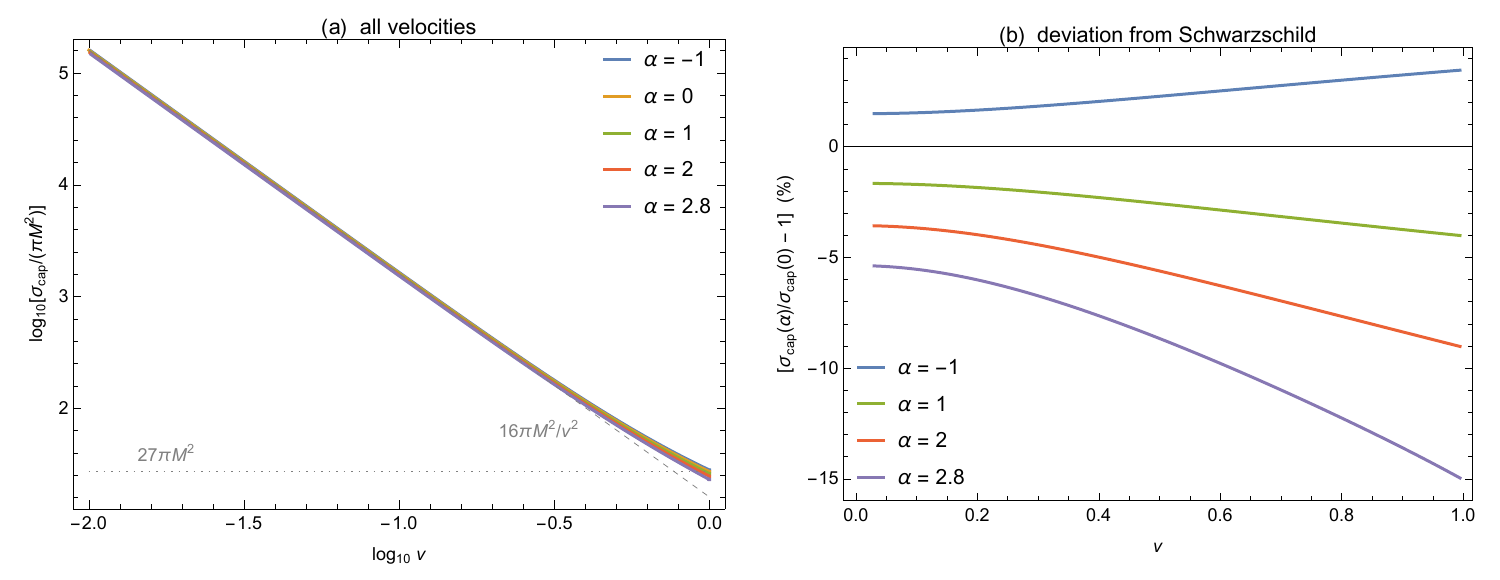}
\caption{Capture cross section of neutral massive particles coming from
infinity, for $b=0$. Left: $\sigma_{\rm cap}$ as a function of the velocity. The
gray lines show the non-relativistic limit $16\pi M^2/v^2$ and the
ultrarelativistic limit $27\pi M^2$. Right: relative deviation from the
Schwarzschild value.}\label{fig:sigmav}
\end{figure*}

The cross section as a function of the velocity is shown in
Fig.~\ref{fig:sigmav}. The deviation from Schwarzschild grows with $v$. For
$\alpha=2.8$ the cross section is $5.4\%$ smaller for $v\to0$, when the particle
turns around far from the black hole and does not feel much the correction
term, and $15.0\%$ smaller for $v\to1$. For $\alpha=-1$ it is instead $1.5\%$
and $3.5\%$ larger. Notice that for $v\to1$ we recover the value of
Table~\ref{tab:capture} for photons, which provides another check of the
numerics. We can then say that a qOS black hole with $\alpha>0$ is at the same
time a more efficient radiator and a smaller absorber than a Schwarzschild black
hole of the same mass, while the opposite is true for $\alpha<0$. Both effects
are due to the repulsive nature of the correction term, which weakens gravity at
small $r$, moves the circular orbits inwards and lowers the potential barrier.

Two remarks are in order. First, for $\alpha>27/16$ the lapse~\eqref{eq:f} has
no zeros. Its minimum is located at $r=(2\alpha M)^{1/3}$, where $f=0.055$ for
$\alpha=2$ and $f=0.155$ for $\alpha=2.8$, and $f\to+\infty$ for $r\to0$. Then a
particle that crosses the potential barrier reaches a minimum radius and comes
back. Therefore for $\alpha>27/16$ Eqs.~\eqref{eq:sigmaph}
and~\eqref{eq:sigmamassive} describe the penetration of the barrier rather than
the absorption of particles, although all the numbers quoted above do not
change. Second, writing the angular momentum in terms of the impact parameter as
$L=mv_\infty\beta$ requires that the particle comes from infinity. As we shall
see below this is not possible for $b\neq0$. Therefore in the magnetized panels
of Figs.~\ref{fig:rmb} and~\ref{fig:capture} the marginally bound orbit and its
angular momentum are well defined, but their interpretation as a cross section
is only formal.

\subsection{Charged particles and magnetic confinement}\label{sec:chargedcapture}

For $b\neq0$ there is no cross section defined at infinity. In fact, from
Eq.~\eqref{eq:Veff} we see that $V_{\rm eff}\simeq b^2r^2/4\to\infty$ 
at large
distances, so that the uniform magnetic field confines charged particles to a
finite region. This is a well known feature of the Wald
field~\cite{FrolovShoom2010,Kolos2015}.

\begin{figure*}[t]\includegraphics[width=0.98\textwidth]{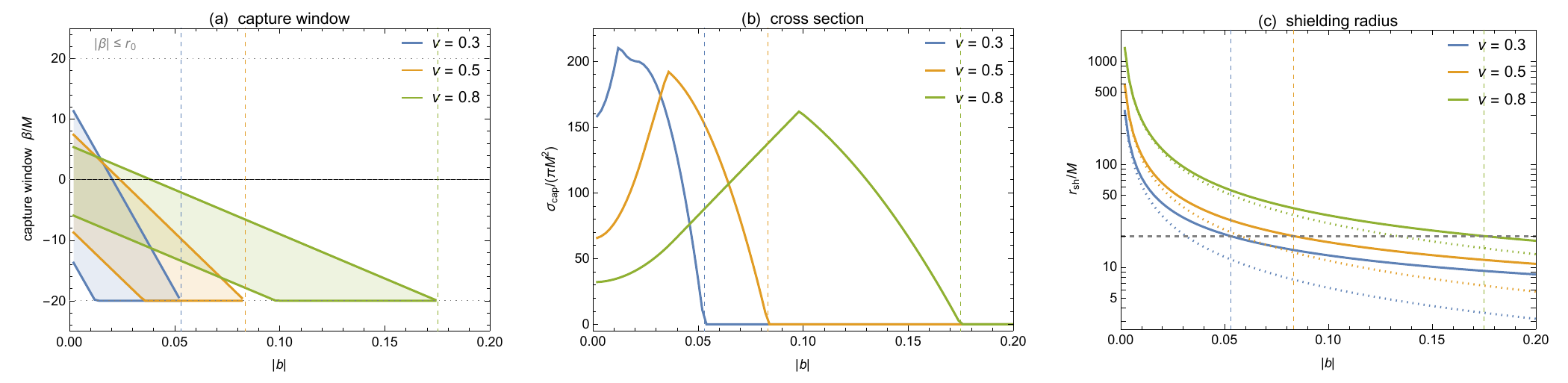}
\caption{Capture of charged particles injected at $r_0=20M$ for $\alpha=0.6$ and
three values of the local velocity $v$. Left: capture window for the signed
local impact parameter. The dotted lines show the bound $|\beta|\le r_0$.
Middle: cross section~\eqref{eq:sigmacharged}. Right: magnetic shielding
radius, exact (solid lines) and approximated by $2\gamma v/|b|$ (dotted lines),
with the horizontal dashed line showing $r_0$. The vertical dashed lines, at
$|b|=0.0529$, $0.0833$ and $0.1749$, show the values at which the capture window
closes, the cross section vanishes and the shielding radius equals $r_0$, which
coincide as expected.}\label{fig:charged}
\end{figure*}

We can then consider particles injected from a sphere of finite radius $r_0$
with locally measured velocity $v$. In the static orthonormal frame the
constants of motion are $E=\gamma\sqrt{f(r_0)}$ and
$L=\gamma v_{\hat\phi}r_0+\tfrac b2 r_0^2$, and we define the signed local impact
parameter as $\beta=r_0\zeta$, with $\zeta=v_{\hat\phi}/v\in[-1,1]$. The particle
is captured if there are no turning points between $r_h$ and $r_0$, i.e. if
\begin{equation}
\gamma^2f(r_0)>\max_{r_h<r<r_0}f(r)\left[1+\frac{\left(L-\tfrac b2r^2\right)^2}{r^2}\right].
\label{eq:capcondcharged}
\end{equation}
The values of $\beta$ that satisfy this condition form an interval
$[\beta_-,\beta_+]$, and we define the cross section as
\begin{equation}
\sigma_{\rm cap}=\frac{\pi}{2}\left[
{\rm sgn}(\beta_+)\beta_+^2-{\rm sgn}(\beta_-)\beta_-^2\right],
\label{eq:sigmacharged}
\end{equation}
which reduces to $\pi\beta_{\rm cr}^2$ in the neutral case, where the interval
is symmetric. Since $L$ contains the term $\tfrac b2 r_0^2$, capture is possible
only if $\tfrac{|b|}{2}r_0^2-\gamma vr_0\lesssim L_{\rm cr}$. This defines a
magnetic shielding radius
\begin{equation}
r_{\rm sh}\simeq\frac{2\gamma v}{|b|},
\label{eq:rshield}
\end{equation}
beyond which a charged particle with velocity $v$ can not reach the black hole,
whatever its direction.

The results are shown in Fig.~\ref{fig:charged}. The left panel shows the
capture window $[\beta_-,\beta_+]$. For $b=0$ it is symmetric, while for $b\neq0$
it moves towards retrograde injection, because the magnetic term
$\tfrac b2 r_0^2$ must be compensated by the kinetic angular momentum for the
particle to reach the black hole. Prograde particles are instead deflected away
and, for example, for $v=0.5$ no particle injected in the prograde direction is
captured for $|b|\gtrsim0.024$. Then the window reaches the bound
$|\beta|\le r_0$ and it closes completely at $|b|\simeq0.083$, when $r_0$
becomes larger than the shielding radius. For faster particles the window
remains open up to larger values of $|b|$.

The cross section~\eqref{eq:sigmacharged}, shown in the middle panel, is not
monotonic. It first grows with $|b|$, since retrograde particles with large
impact parameters are focused by the Lorentz force, and then it goes to zero
when the window reaches the geometric bound. The right panel shows that the
exact shielding radius follows closely the estimate~\eqref{eq:rshield}, with a
small offset due to $L_{\rm cr}\sim4M$. For realistic values of the coupling this
radius is very large compared with the size of the black hole. Then the
magnetized qOS black hole is effectively screened from the charged particles in
its environment, which can only be accreted through the disk, where viscosity
removes their angular momentum. This provides a justification for the disk model
used in Sec.~\ref{sec:disk}.

\section{Discussion}\label{sec:concl}

We studied the observational properties of the quantum Oppenheimer--Snyder black
hole immersed in an external uniform magnetic field. We considered a
Novikov--Thorne accretion disk made of weakly charged matter, the iron K$\alpha$
line emitted by the disk and the capture cross sections for massless, massive
and charged particles. We considered the whole range of the quantum parameter
for which a thin disk exists. This range is not bounded by the extremal value
$\alpha=27/16$, above which the object is horizonless, but by
$\alpha_{\max}=729/256$, where the photon sphere disappears. In fact, while the
ISCO exists up to $\alpha=5.353234$, for $\alpha>\alpha_{\max}$
the circular orbits extend all the way to the static ring $r=(2\alpha M)^{1/3}$
and the inner boundary condition of the disk loses its meaning. Since the
geometry is static, the Wald potential is purely axial and the field affects
the disk only through the Lorentz force acting on the charged matter.

Our main result is that the two parameters affect the observables in very
different ways. Over the whole range of $\alpha$ the efficiency of the disk
changes by less than half a percentage point, the maximum of the flux by $32\%$
and the position of the peak of the spectrum by $0.025$~dex. On the other hand,
the magnetic coupling increases the efficiency up to $21.7\%$, the maximum of
the flux by a factor of about $18$ and the peak of the spectrum by a factor of
$4$, moving it to frequencies about twice larger. The iron K$\alpha$ line
shows the same hierarchy. The quantum parameter only moves the red edge of the
line by a few percent, as it happens for the asymptotically safe black hole of
Ref.~\cite{Urmanov2025}, which has the same lapse at first order. The magnetic
coupling, instead, reduces the orbital velocity of the gas and suppresses the
Doppler blue peak, so that the line becomes a single peak below the rest frame
energy.

For the capture of particles the situation is reversed. Since photons do not
feel the magnetic field, $\sigma_{\rm ph}$ depends only on $\alpha$ and it
decreases from $1.035$ to $0.850$ in units of $27\pi M^2$. The cross section
for massive particles also decreases with $\alpha$, by up to $15\%$ in the
ultrarelativistic limit. Therefore a qOS black hole with $\alpha>0$ is at the
same time a better radiator and a smaller absorber than the Schwarzschild black
hole with the same mass. Charged particles, instead, can not reach the black
hole from beyond the shielding radius $r_{\rm sh}\simeq2\gamma v/|b|$, and the
capture is strongly asymmetric between prograde and retrograde particles.

These results suggest that, in principle, the two parameters could be
distinguished by combining observations of the disk spectrum, which are
dominated by $b$, with observations of the shadow~\cite{EHT2019}, which within
the present test field model depend only on $\alpha$. It is important to
remark, however, that the effects of $\alpha$ are small and comparable with the
uncertainties on the mass, the spin and the emissivity profile of real sources.
Also, our analysis is limited to the static case, while astrophysical black
holes are expected to rotate. For a rotating generalization of the qOS metric
the Wald potential would have an electrostatic component and the disk would no
longer be symmetric under $b\to-b$. As an outlook, it would be interesting to
consider rotating models and ray traced images of the disk, and to compare the
results with the case of black holes surrounded by dark matter halos.

\begin{acknowledgments}
AI acknowledges financial support from SECIHTI through the National Postgraduate
Scholarship Program (CVU 2222058). This work was supported by DGAPA-PAPIIT UNAM, Grant No. 108225, and
Conahcyt, grant No. CBF-2025-I-243.
\end{acknowledgments}

\appendix

\section{Expansions for small $\alpha$}\label{app:series}

When the quantum correction is small, i.e. $|\alpha|\ll M^2$, the characteristic
radii and the capture cross sections can be written as power series in $\alpha$
around their Schwarzschild values. Here we report them up to third order.

\subsection{Characteristic radii}

The outer horizon, the photon sphere, the marginally bound orbit and the ISCO
(for $b=0$) are simple roots of the polynomial
\begin{equation}
r^4-2Mr^3+\alpha M^2=0
\label{eq:horpoly}
\end{equation}
and of Eqs.~\eqref{eq:photonsphere}, \eqref{eq:mbpoly}
and~\eqref{eq:iscopoly}, respectively, which reduce to $r_0=2M$, $3M$, $4M$ and
$6M$ for $\alpha=0$. Writing each of these equations as $P(r,\alpha)=0$ and
substituting $r=r_0+c_1\alpha+c_2\alpha^2+c_3\alpha^3$, at each order in $\alpha$
we obtain a linear equation for the corresponding coefficient,
\begin{equation}
P_{,r}(r_0,0)\,c_n+S_n(c_1,\dots,c_{n-1})=0 ,
\end{equation}
where $S_n$ depends only on the coefficients already determined. In particular
$c_1=-P_{,\alpha}/P_{,r}$, evaluated at $r=r_0$ and $\alpha=0$. For example, for
the photon sphere we can write $r=3M+\delta$, so that
Eq.~\eqref{eq:photonsphere} becomes $(3M+\delta)^3\delta=-3\alpha M^2$, and the
first three orders give
\begin{align}
&27M^3c_1=-3M^2,\qquad 27M^3c_2+27M^2c_1^2=0,\nonumber\\
&27M^3c_3+54M^2c_1c_2+9Mc_1^3=0 .
\end{align}
Proceeding in the same way for the other equations we obtain, up to terms of
order $\alpha^4$,
\begin{align}
r_h&=2M-\frac{\alpha}{8M}-\frac{3\alpha^2}{128M^3}-\frac{15\alpha^3}{2048M^5},\label{eq:rhser}\\
r_{\rm ph}&=3M-\frac{\alpha}{9M}-\frac{\alpha^2}{81M^3}-\frac{5\alpha^3}{2187M^5},\label{eq:rphser}\\
r_{\rm mb}&=4M-\frac{\alpha}{8M}-\frac{39\alpha^2}{4096M^3}-\frac{327\alpha^3}{262144M^5},\label{eq:rmbser}\\
r_{\rm ISCO}&=6M-\frac{11\alpha}{72M}-\frac{89\alpha^2}{10368M^3}-\frac{3743\alpha^3}{4478976M^5}.\label{eq:riscoser}
\end{align}
All the coefficients are negative, so that all the radii decrease for
$\alpha>0$. At first order the relative shifts are $-\alpha/(16M^2)$,
$-\alpha/(27M^2)$, $-\alpha/(32M^2)$ and $-11\alpha/(432M^2)$, i.e. the
correction is larger for the radii closer to the center, as discussed in
Sec.~\ref{sec:mb}. It is worth noticing that for the horizon and the photon
sphere the series can be obtained to all orders by Lagrange inversion, which
gives
\begin{align}
r_h&=2M\Big[1-\sum_{n\ge1}C_n\Big(\frac{\alpha}{16M^2}\Big)^n\Big],\nonumber\\
r_{\rm ph}&=3M\Big[1-\sum_{n\ge1}C_n\Big(\frac{\alpha}{27M^2}\Big)^n\Big],
\end{align}
with $C_n=(4n-2)!/[n!(3n-1)!]$. These series converge for
$|\alpha|<27M^2/16$ and $|\alpha|<729M^2/256$, respectively, i.e. exactly up to
the critical values~\eqref{eq:alphahor} and~\eqref{eq:alphamax}.

\subsection{Critical impact parameter and capture cross sections}

From the capture conditions for photons and for massive particles with $E=1$,
the critical impact parameter and the angular momentum of the marginally bound
orbit are minima,
\begin{equation}
b_{\rm cr}^2=\min_r\frac{r^2}{f},\qquad
L_{\rm mb}^2=\min_r\frac{r^2(1-f)}{f} .
\end{equation}
Since the derivative with respect to $r$ vanishes at the minimum, the
derivatives with respect to $\alpha$ come only from the explicit dependence of
$f$, with $\partial_\alpha f=M^2/r^4$, and are given exactly by
\begin{equation}
\frac{db_{\rm cr}^2}{d\alpha}=-\frac{M^2}{r_{\rm ph}^2f_{\rm ph}^2},\qquad
\frac{dL_{\rm mb}^2}{d\alpha}=-\frac{M^2}{r_{\rm mb}^2f_{\rm mb}^2},
\end{equation}
where $f_{\rm ph}=f(r_{\rm ph})$ and $f_{\rm mb}=f(r_{\rm mb})$. At $\alpha=0$
these are equal to $-1$ and $-1/4$ and give the first order terms. The higher
orders follow by inserting Eqs.~\eqref{eq:rphser} and~\eqref{eq:rmbser} into the
exact relations
\begin{equation}
b_{\rm cr}^2=\frac{3r_{\rm ph}^3}{2r_{\rm ph}-3M},\qquad
L_{\rm mb}^2=\frac{r_{\rm mb}^2\left(1-f_{\rm mb}\right)}{f_{\rm mb}},
\end{equation}
where the first follows from Eq.~\eqref{eq:photonsphere} and the second from the
condition $2f^2-2f+rf'=0$. Up to terms of order $\alpha^4$ we obtain
\begin{align}
b_{\rm cr}&=3\sqrt3M-\frac{\sqrt3\,\alpha}{18M}-\frac{\sqrt3\,\alpha^2}{216M^3}-\frac{73\sqrt3\,\alpha^3}{104976M^5},\\
L_{\rm mb}&=4M-\frac{\alpha}{32M}-\frac{15\alpha^2}{8192M^3}-\frac{201\alpha^3}{1048576M^5},
\end{align}
and the cross sections
\begin{align}
\sigma_{\rm ph}&=\pi\left(27M^2-\alpha-\frac{2\alpha^2}{27M^2}-\frac{8\alpha^3}{729M^4}\right),\\
\sigma_{\rm capt}&=\frac{\pi}{v^2}\left(16M^2-\frac{\alpha}{4}-\frac{7\alpha^2}{512M^2}-\frac{93\alpha^3}{65536M^4}\right).
\end{align}

The same argument applies to massive particles with arbitrary velocity, for
which $L_{\rm cr}^2=\min_r r^2(\gamma^2/f-1)$. At first order in $\alpha$ we find
\begin{equation}
\frac{\sigma_{\rm cap}(\alpha,v)}{\sigma_{\rm cap}(0,v)}=1-\frac{\gamma^2M^2\,\alpha}{r_c^4f_c\left(\gamma^2-f_c\right)}+O(\alpha^2),
\label{eq:sigmavser}
\end{equation}
where
\begin{equation}
r_c=M\,\frac{3\gamma^2-4+\gamma\sqrt{9\gamma^2-8}}{2\gamma^2v^2}
\end{equation}
is the radius of the unstable circular orbit of the Schwarzschild metric with
energy $\gamma$ and $f_c=1-2M/r_c$. For $v\to0$ and $v\to1$ we have $r_c\to4M$ and
$r_c\to3M$, and Eq.~\eqref{eq:sigmavser} reproduces the first order terms of
$\sigma_{\rm capt}$ and $\sigma_{\rm ph}$, respectively.

Finally, the largest value of the magnetic coupling for which the marginally
bound orbit exists, Eq.~\eqref{eq:bmax}, becomes, to the same order,
\begin{align}
b_{\max}={}&\frac{\sqrt2}{6M}+\frac{\sqrt2\,\alpha}{648M^3}+\frac{11\sqrt2\,\alpha^2}{139968M^5}\nonumber\\
&+\frac{29\sqrt2\,\alpha^3}{15116544M^7}.
\end{align}

\bibliographystyle{apsrev4-2}
\bibliography{0refs}

\end{document}